\documentclass[lettersize,journal]{IEEEtran}
\usepackage{amsmath,amsfonts}
\usepackage{algorithmic}
\usepackage{algorithm}
\usepackage{array}
\usepackage[caption=false,font=normalsize,labelfont=sf,textfont=sf]{subfig}
\usepackage{textcomp}
\usepackage{stfloats}
\usepackage{url}
\newcolumntype{?}[1]{!{\vrule width #1}}
\usepackage{booktabs}
\usepackage{verbatim}
\usepackage{graphicx}
\usepackage{caption}
\usepackage{cite}
\begin{document}

\title{Vertiport Design Methodology and Capacity Analysis}
% \title{Queueing Theory Approach to Trade-offs in Vertiport Design Parameters}

\author{Emin Burak Onat, Jean-Luc Lupien, and Raja Sengupta
        % <-this % stops a space
\thanks{Emin Burak Onat, Jean-Luc Lupien and Raja Sengupta are with the Civil and Environmental Engineering Department, University of California at Berkeley, Berkeley, CA 94720 USA. This research was funded by Supernal, grant number 052838-002}% <-this % stops a space
}



\maketitle

\begin{abstract}
This paper presents a comprehensive methodology for vertiport design and capacity analysis using stochastic queueing models. Vertiport capacity is defined as the maximum steady-state throughput of aircraft per hour while maintaining a specified maximum queuing delay with high probability. We introduce a two-node queueing network model with a finite buffer to analytically derive mean queue lengths and delays at single-TLOF vertiports, supported by a simulation model that validates these findings under various operational conditions. Our analyses reveal that vertiport capacity is significantly constrained by the interactions between takeoff/landing operations and turnaround activities. Capacity is further reduced by the variance in these operational parameters, with deterministic times substantially enhancing capacity. The study shows that vertiport throughput is typically around 50\% of the theoretical maximum departure capacity in a Markovian model, decreasing sharply as confidence in maximum queuing delay increases. Non-Markovian models with adequate parking pads achieve full TLOF capacity. Additionally, our methodology identifies the optimal number of parking pads required to maximize throughput per unit area. The findings provide insights for designing vertiport infrastructure to support scalable and efficient UAM deployment.
\end{abstract}


% TODOs
% Airport literature, why you're different
% Write the 1,2,3.rd order results
% Write the outline of the paper
% Why M/M/1 with lambda = 30 and M/M/2 with lambda = 60 variance are different?

\section{Introduction}
\IEEEPARstart{T}{he} emergence of Urban Air Mobility (UAM) holds the potential to introduce a new mode of urban transportation, with electric vertical take-off and landing (eVTOL) aircraft promising efficient and sustainable mobility \cite{drones7050319}. As the backbone of the UAM ecosystem, vertiports are hubs for the takeoff, landing, and servicing of eVTOL aircraft. A thorough understanding of the operational parameters and design trade-offs of these hubs is crucial for ensuring the operational viability and scalability of UAM. 

Specifically, determining the \emph{capacity} of a UAM vertiport is of primary design importance. We define capacity as the maximum steady-state throughput of aircraft per hour given that the maximum queuing delay for holding aircraft remains below a specified threshold, $W_q$, with high probability. Capacity is derived as a function of structural and operational parameters. Structural design parameters include the number of parking pads, while operational parameters encompass the number of takeoff and landing operations per hour at a Touch-Down and Lift-Off area (TLOF) and the rate at which aircraft are charged and turned around.

We propose that quantifying vertiport capacity presents a distinct and more complex challenge compared to conventional airports. Vertiports, designed for high-density urban areas, face significant limitations in footprint area, resulting in fewer gates compared to traditional airports. Servicing eVTOL aircraft adds to this complexity due to the inherent uncertainty of the charging process and the fact that charging and takeoff/landing times are of the same order of magnitude.  \cite{onat2024simulationoptimization}
indicate that a typical 30-minute UAM flight necessitates approximately 6 minutes of charging to restore the 160kWh battery powered aircraft's pre-flight state of charge (SoC) when it is at a lower SoC levels, while takeoff and landing operations each take about 1 minute. This similarity in the duration of charging and takeoff/landing times, which differs from conventional airport operations, leads to a highly interactive and complex stochastic system that must be analyzed as a whole.

This paper advances our understanding of vertiports by using stochastic queueing models to quantify the capacity of a UAM vertiport. We present a two-node queueing network model with a finite buffer and an analytical way to calculate the mean queue length and mean waiting time while obtaining confidence intervals for these values. This model assumes that aircraft take off or land at the TLOF and conduct all other operations, including charging, at the parking pads. Logically, the maximum throughput of a vertiport must be smaller than the minimum of the takeoff/landing, and turnaround capacity. However, our analyses show vertiport capacity to be significantly smaller than these bounds, when the queueing delay of the aircraft are required to be small with high probability. 

Complementing this, our simulator replicates the relationships identified by the analytical model. The analysis reveals that reductions in capacity, necessary to maintain short queueing delays with high probability, correlate with the variance in TLOF and turnaround performance. More deterministic takeoff/landing and turnaround times can significantly enhance vertiport capacity, even under strict delay constraints. Furthermore, our methodology identifies the optimal number of parking pads at a vertiport to maximize throughput gain per unit area.

\subsection{Literature Review}
The capacity problem has been studied by using both deterministic and stochastic methods. Deterministic analyses include \cite{vascik2019development} and \cite{gimpo_airport} who developed an integer programming approach to analytically estimate vertiport capacity envelopes. The sensitivity of vertiport capacity to various topological and operational factors was then examined. \cite{guerreiro2020capacity} conducted a deterministic queueing analysis to understand the capacity of vertiports, using simulations. However, deterministic analysis does not consider the inherently stochastic nature of vertiport operations. Therefore, the obtained values can only be seen as upper-bounds to the real-world capacity of a vertiport. Additionally, these studies disregard the aircraft queueing delay in the airspace which could significantly affect vertiport capacity.

Stochastic queueing analysis is also applied in the study of vertiports, drawing on methodologies established in airport research. Single-node queueing models, which represent a runway, have been used extensively to estimate runway delays. These models typically assume an infinite buffer capacity, analogous to an unlimited number of gates. Such a model is particularly suited to airports, where the primary capacity constraint lies with the runways, due to the relative abundance of gates compared to runways \cite{Newell1979, Malone1995, Kivestu1976, simaiakis2016queuing, Itoh2022}. 

However, vertiports are distinct from airports in that they are more likely to be constrained by gate availability rather than by TLOF (equivalent to runway) operations. This reality was first underlined by \cite{kim1995vertiport} where the authors developed a single-node queueing model to estimate delays during periods of steady demand. Recently, \cite{zhang2022method} modeled vertiport operations by separating them into three distinct phases: arrival, servicing, and departure. Each process was represented by an independent queue with different service rates. Capacity is then determined from whichever process is considered limiting. 

These analyses fall short in a key aspect. As explored in this paper, the interaction of these stochastic processes create a queuing network structure. A proper analysis, therefore, cannot consider queues individually. Rather, the entire stochastic system must be modeled including all interactions between queues.

% vertiport capacity is a certain proportion of TLOF capacity. This proportion is influenced by the variance in operational parameters and can be significantly enhanced by minimizing uncertainties in both the TLOF service rate and turnaround rates. 

The organization of the rest of the paper is outlined as follows: The vertiport queuing model is delineated in Section \ref{vertiport_ops_model}. A concise description of the simulation model is offered in Section \ref{sim_model}. Comparative results from both the analytical and simulation models are presented in Section \ref{results}. Finally, Section \ref{conclusions} draws conclusions from the study's findings.

\section{Vertiport Queueing Network Model} \label{vertiport_ops_model}
Since their formal introduction by Erlang in \cite{erlang}, queueing theory has been applied to numerous engineering applications. In particular, queueing systems, where multiple queues interact with each other, are great models for telecommunications, logistics, finance, and transportation \cite{QueueEx1}. A critical aspect of queueing theory is the concept of blocking mechanisms, which is essential for situations where one service process cannot start without the completion of another. Such queueing systems with blocking are studied in \cite{Balsamo2001}. While some well-known queueing systems have closed-form expressions for their state probabilities, queueing systems with blocking often require numerical methods to solve \cite{balsamoreview}.

% Queueing networks are categorized based on buffer capacities into two types: those with infinite buffers and those with finite buffers. 
Networks featuring a queue with a finite buffer are identified as queueing networks with blocking. This classification is particularly relevant for vertiports, where the limited number of parking pads necessitates a blocking mechanism. In such models, reaching a queue's maximum capacity temporarily halts the entry of new entities into the service center until space becomes available, impacting arrivals from both within the network and external sources. 

% In this paper, we model a vertiport system as a two-node open queueing network with blocking and establish relationships between design parameters and vertiport throughput. The representational fidelity of our queueing models is also benchmarked by simulations. 

\begin{figure}
\centering
\includegraphics[width=0.45\textwidth]{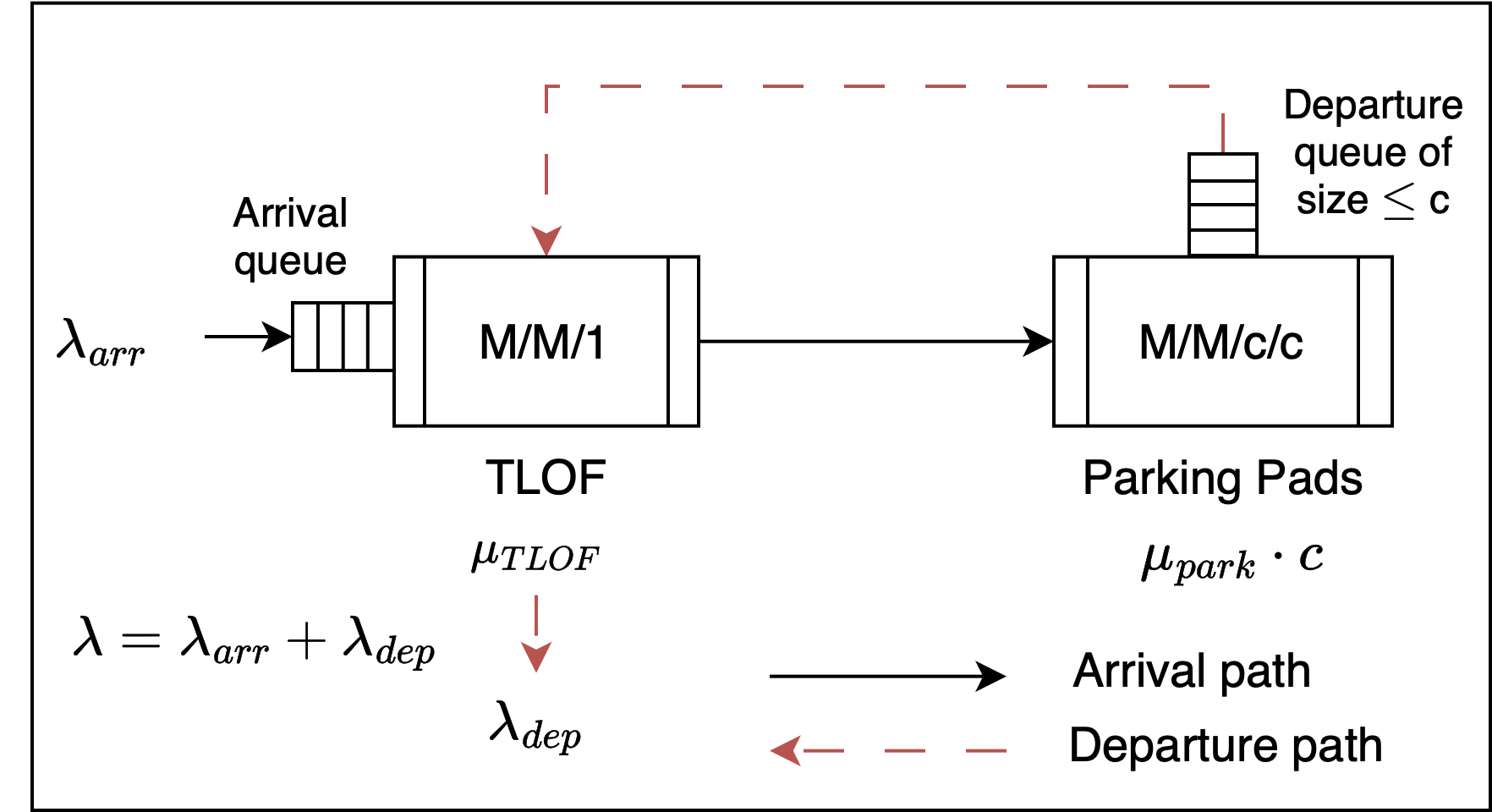}

\caption{Queueing model of a vertiport. The system consists of two main queues: the M/M/1 queue representing the TLOF and the M/M/c/c queue representing the parking pads. Aircraft arrive at the system with a rate of $\lambda_{arr}$ and depart after being serviced with a rate of $\lambda_{dep}$.}
\label{fig:feedback_model}
\end{figure}

The queueing network model for a single-TLOF vertiport, illustrated in Figure \ref{fig:feedback_model}, comprises an arrival queue, a servicing stage represented by the TLOF, and several parking pads where aircraft undergo servicing prior to departure. Initially, an arriving aircraft joins the arrival queue within the terminal airspace. Upon advancing to the head of this queue, the aircraft is processed at the TLOF, which functions as the entry and exit point to the vertiport. After being serviced at the TLOF, the aircraft moves to the parking pads for charging or loading/unloading. Once the service at the parking pad is complete, the aircraft get on the departure queue of TLOF to return to the TLOF for takeoff and exits the system. If the parking pads are full, the TLOF becomes blocked under the Blocking Before Service-Server Not Occupied (BBS-SNO) discipline \cite{Balsamo2001}. This means that no arriving aircraft can be serviced at or occupy the TLOF until a parking pad becomes available. 

In blocking before service (BBS) mechanisms, when an entity attempts to initiate service at node $i$ but finds its target node $j$ at capacity, the service at node $i$ is delayed, effectively blocking this server. Service resumes only when node $j$ has a departure. This blocking is categorized into BBS-SO, where the server can hold an entity during the blockage (server occupied), and BBS-SNO, where the server cannot hold an entity when blocked. BBS-SNO is not well defined except for special type of queueing networks. The operational mechanics of different types of blocking are elaborated in \cite{Balsamo2001}.

We follow Kendall's notation to classify the queues. The TLOF is defined as an M/M/1 queue, where the "M" stands for Markovian, indicating that both the arrival rate and service rate follow an exponential distribution characterized by being memoryless. The third element of the acronym, "1", denotes the presence of a single server that services one entity at a time, and, theoretically, this class of queues allow for an infinite number of entities to wait in the queue. The arrival queue represents the terminal airspace where aircraft hold to be serviced by the TLOF with a service rate of $\mu_{TLOF}$ and "1" represents a one-TLOF vertiport.

The second node in our system is an M/M/c/c queue with 'c' servers where arrivals also follow a Poisson process and service times are exponentially distributed. However, unlike the M/M/1 queue, there is no room for waiting; the system capacity is equal to the number of servers 'c'. This means if all servers are busy, new arrivals are blocked and cleared (i.e., they leave without being serviced). The service rate $\mu_{surface}$ is the total service rate of all parking pads combined, which is $\mu_{park} \cdot c$ where $\mu_{park}$ is the service rate of a single parking pad. The mean time to turnaround an aircraft, $1/\mu_{park}$, encompasses the time to taxi to and from a parking pad, the charging time, and any concurrent turnaround activities. For a vertiport system to maintain a steady state, the arrival rate, $\lambda_{arr}$, must match the departure rate, $\lambda_{dep}$ ($\lambda_{arr} = \lambda_{dep}$). 

This model effectively captures the key dynamics of a vertiport using straightforward parameters. These parameters can be derived from various sources such as design codes, field experiments, analytical studies, and simulations. For instance, runway capacity profiles, empirically established and detailed in FAA reports, provide a basis that can be adapted to evaluate TLOF capacity \cite{faa_airport_profiles}. The charging times, represented by the reciprocal of $\mu_{charge}$, calculated depend on the aircraft's previous flight distance, the next scheduled flight distance, the current state of charge (SoC) \cite{onat2024simulationoptimization}. Given that we model vertiport operations as Continuous Time Markov Chains (CTMCs), which inherently assume an exponential distribution for the timing of events, we can estimate parameters such as $\mu_{TLOF}$ and $\mu_{charge}$ by fitting empirical or simulation data to an exponential distribution. Furthermore, the number of parking pads ($c$) is typically influenced by the available capital investment, land availability, and design specifications set by the FAA \cite{UAMCapacity, faa_brief}. Alternatively, $c$ can be determined based on the desired throughput levels for the vertiport, taking into account the aforementioned constraints.

% Additionally, analytical and simulation studies have been conducted to determine runway capacity \cite{Newell1979, runway_capacity}.

\begin{figure}[ht]
\centering
\includegraphics[width=0.45\textwidth]{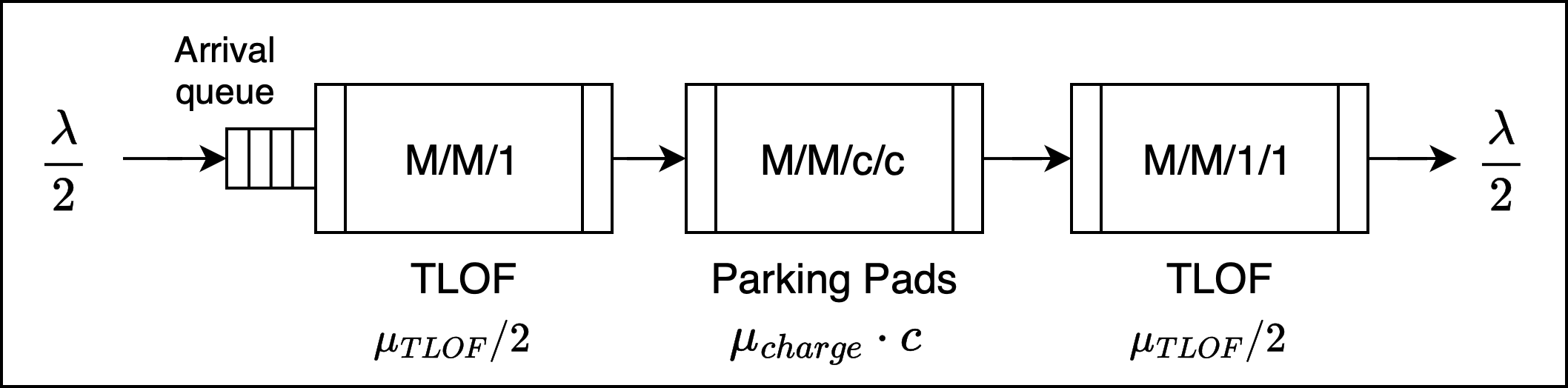}
\caption{Three-node tandem queueing network model}
\label{fig:tandem3node}
\end{figure}

The model shown in Figure \ref{fig:feedback_model} involves three state variables—the number of aircraft in the airspace buffer, in service at parking pads, and in the departure queue—and includes a feedback loop, adding considerable complexity to the system. The combined total of aircraft in service at the parking pads and those in the departure queue cannot exceed $c$, as aircraft in the departure queue must also be accommodated at the parking pads. We refer to this as the feedback model. To simplify the analysis and eliminate the feedback mechanism, we reconfigure the system into a tandem network, as illustrated in Figure \ref{fig:tandem3node}. In the simplified model, we adjust the service rates of the M/M/1 queues by halving them to mirror the service dynamics of the feedback system, which is necessary due to the doubling of TLOFs in the reconfigured model. We also transform the second M/M/1 queue into an M/M/1/1 queue to reflect its transition to a BBS-SO queue discipline. In this case, TLOF server can be occupied since we decoupled the arrivals and departures.

\begin{figure}[ht]
\centering
\includegraphics[width=0.45\textwidth]{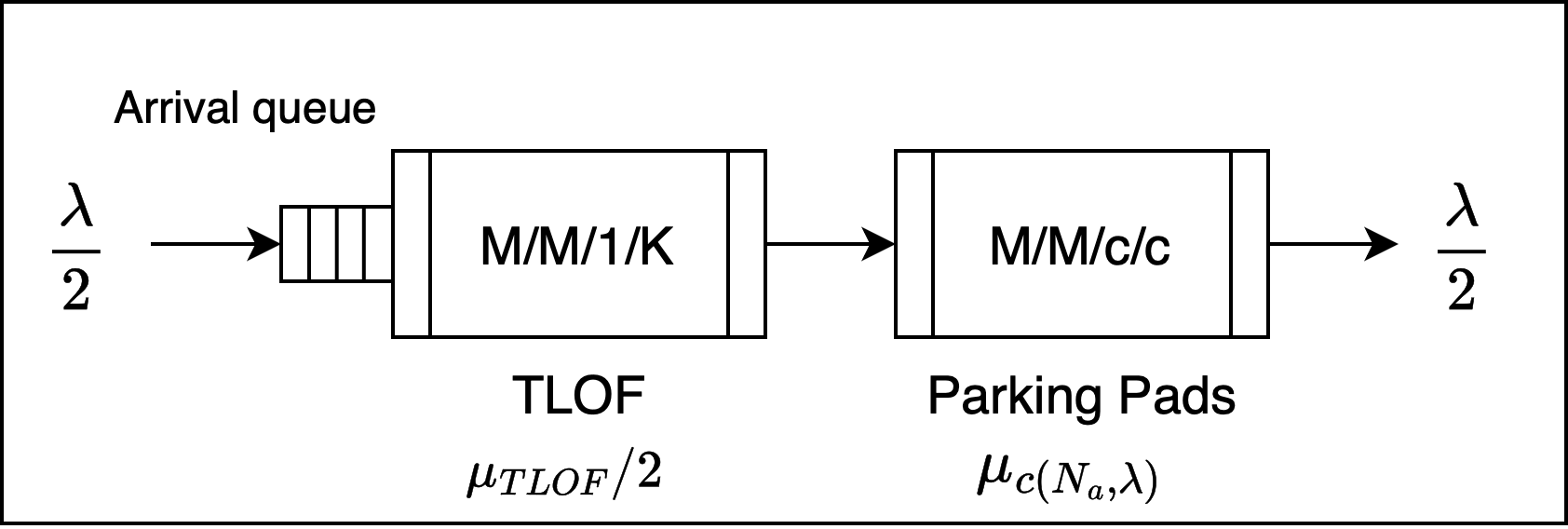}
\caption{Two-node tandem queueing network model}
\label{fig:simplified_model}
\end{figure}

To further streamline the analysis of the vertiport system, we consolidate the second and third queues by merging their service rates. This adjustment transforms the vertiport into a 2-node tandem queue model with finite buffer capacities at both queues, as shown in Figure \ref{fig:simplified_model}. This model features two principal queues: an M/M/1/K queue for the TLOF, characterized by an arrival rate, $\lambda$, and a service rate, $\mu_{TLOF}$, and an M/M/c/c queue for the parking pads with a combined service rate, $\mu_c(N_a)$. The final network utilizes the BBS-SNO blocking mechanism because the initial problem prohibits the occupation of the TLOF server. We set K to a high number, 100, to ensure minimal impact on system dynamics while maintaining analytical tractability with a finite state space for solving global balance equations. The service rate $\mu_c(N_a, \lambda)$ represents the cumulative effect of parking pad servicing and TLOF service for departure, as a function of the number of aircraft on the ground, $N_a$, and the arrival rate, $\lambda$.  The formulation of $\mu_c(N_a, \lambda)$ is provided in equation \ref{eqn:mu_c}. Through this simplification, we reduce the number of state variables and enable the computation of state probabilities using global balance equations. This refined two-node queueing network model approximates the feedback model shown in Fig. \ref{fig:feedback_model} using two TLOFs (one for arrival, one for departure). We will further investigate the accuracy of this approximation in section \ref{results}. In the next section we provide global balance equations that models the vertiport.

\subsection{Notation}
The state of the system is represented by a pair \((i, j)\), where:
\begin{itemize}
    \item \(i\) denotes the number of aircraft waiting in the airspace buffer ($i \leq K$).
    \item \(j\) denotes the number of aircraft at the surface (i.e., charging at a parking pad) ($j \leq c$).
\end{itemize}

The key parameters of the model are defined as follows:
\begin{itemize}
    \item \(\lambda\): Arrival rate of aircraft into the airspace buffer.
    \item \(\mu_{charge}\): Charge rate of the single charger at a parking pad.
    \item \(\mu_{\text{TLOF}}\): Service rate at the TLOF.
    \item \(c\): Number of parking pads available at the vertiport surface.
    \item \(K\): Maximum size of the airspace buffer.  
    \item \(p(i, j)\): Probability of the system being in state \((i, j)\).
    \item $E$: Set of all possible states.
\end{itemize}

\subsection{State Transition Rate Diagram}

\begin{figure}[ht]
\centering
\includegraphics[width=0.49\textwidth]{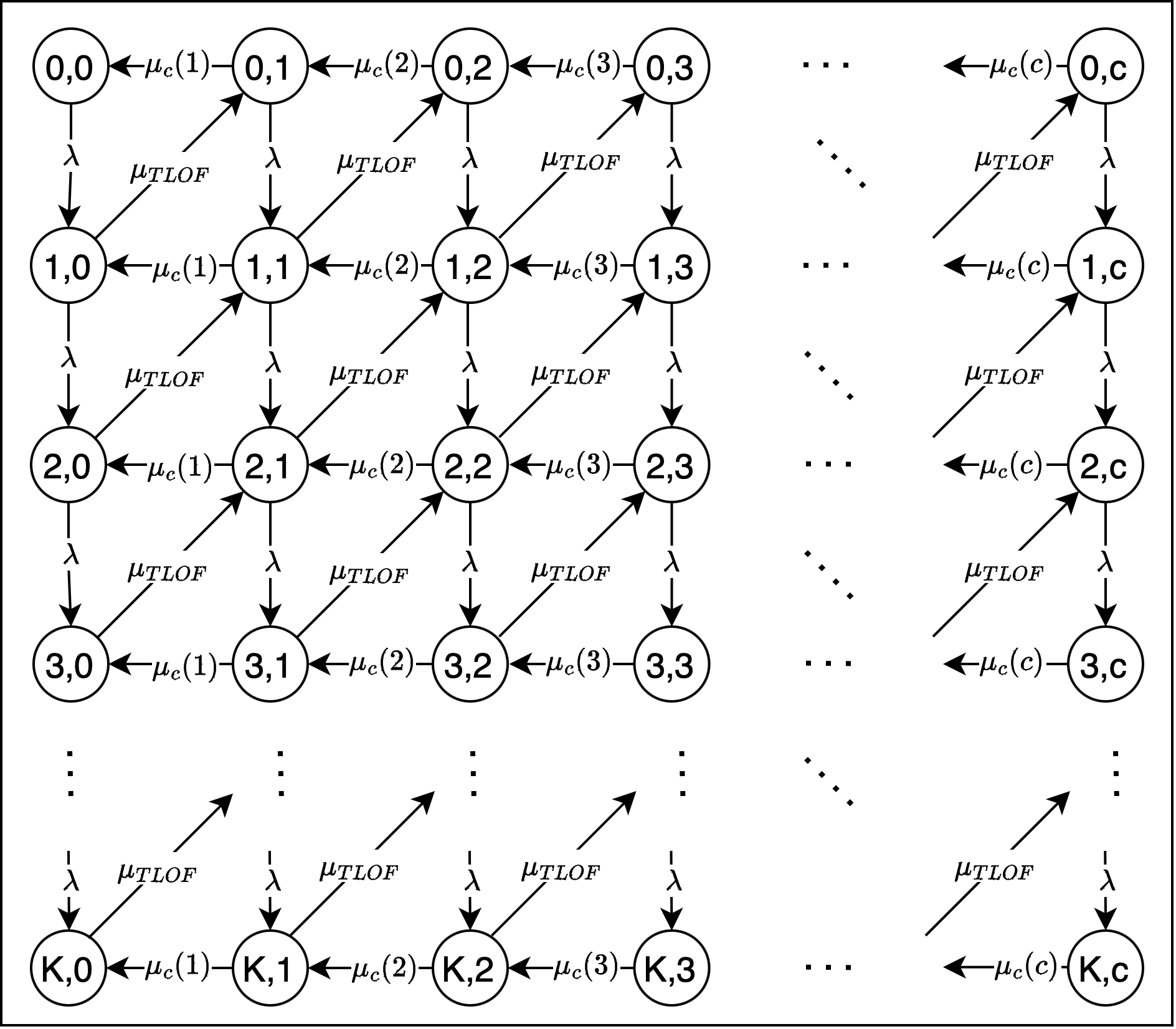}
\caption{State transition diagram for the vertiport queueing system.}
\label{fig:state_transition_diagram}
\end{figure}

The state transition diagram depicted in Figure \ref{fig:state_transition_diagram} illustrates dynamics of a vertiport system. Each circle represents a different state of the system, labeled by the pair \((i, j)\). The transitions between states are governed by transition rates:
\begin{itemize}
    \item Arrows labeled with \(\lambda\) represent aircraft arrivals to the buffer.
    \item Arrows labeled with \(\mu_{\text{TLOF}}\) represent transition rates from the buffer to a parking pad.
    \item Arrows labeled with \(\mu_c\) represent the departure rate of aircraft from the parking pads following charging. The rate \(\mu_c\) is an adjusted charging rate, distinct from \(\mu_{\text{charge}}\), to incorporate the subsequent TLOF service. This adjustment is necessary because, in a vertiport with a single TLOF, an aircraft must utilize the TLOF for departure before the next aircraft can land and occupy the vacated parking pad. 
    
    %The adjusted charge rate is computed as follows: 
    %\begin{equation} \label{eqn:mu_c}\mu_{c}(N_a) = \frac{1}{\frac{1}{\mu_{charge}*N_a} + \frac{1}{\mu_{TLOF}}} \end{equation}

The effective service rate $\mu_c$ can be calculated by considering the combined two-server system composed of the parking pads and departure TLOF. In the two-node tandem queue model, since the charge rate is exponentially distributed and there are no blocking constraints for departures, the arrivals at the TLOF on the departure side adhere to a Poisson process with parameter $\lambda$. This assumption holds because, for systems that are stationary, i.e., have finite queue lengths, the arrival process to any queue in the network remains a Poisson process. Therefore, we have that the average time before the departure TLOF is free can be expressed in the following way:
\[t_{free} = p_0\times 0 + p_1\times \mu_{\text{TLOF}},\]
where $p_0$ and $p_1$ are the probabilities of having 0 and 1 aircraft, respectively, in the queue. These probabilities are given by $p_0 = 1-\frac{\lambda}{\mu_{\text{TLOF}}}$ and $p_1 = \frac{\lambda}{\mu_{\text{TLOF}}}$ from the analysis of \cite{balsamoBook}. This gives the average wait time for a free departure TLOF $t_{free} = \frac{\lambda}{\mu_{\text{TLOF}}^2}$. Finally, the service rate of the server is given by:

\begin{equation} \label{eqn:mu_c}
\mu_c(N_a, \lambda) = \frac{1}{\frac{1}{\mu_{\text{charge}}*N_a}+\frac{\lambda}{\mu_{\text{TLOF}}^2}}
\end{equation}

\end{itemize}

The diagram also shows the terminal airspace capacity (along the vertical axis) and the number of parking pads (along the horizontal axis), which constrain the possible state space.

\subsection{Exact Analysis of Markovian Process}
Under the assumptions of exponential service times at charging pads and the independence between these service times and aircraft arrival times, the evolution of the vertiport system can be described by a continuous-time ergodic Markov chain with a discrete state space $E$ and a transition rate matrix $Q$. The steady-state and transient behavior of the vertiport can thus be studied through the Markov process framework. Assuming an irreducible network routing matrix, which reflects a vertiport system where every state is reachable from any other state, there exists a unique steady-state probability distribution over the network's states, denoted by $\pi = [\pi(i,j)]$ for all $(i,j) \in E$.

This steady-state probability distribution $\pi$ can be found by solving the global balance equations given by the homogeneous linear system $\pi Q = \mathbf{0}$, where $\mathbf{0}$ is the zero vector. The solution must also satisfy the normalization condition $\sum_{(i,j) \in E} \pi(i,j) = 1$.

The global balance equations for the vertiport system are presented below:

\begin{flalign} \label{2nd}
    & \begin{aligned}
        & \mu_c(j+1) \cdot p[i, j+1] - \lambda \cdot p[i,j] = 0; \\
        & \quad \text{for } i=0, j=0
    \end{aligned} &
\end{flalign}

\begin{flalign}  \label{3rd}
    & \begin{aligned}
        & \lambda \cdot p[i-1, j] + \mu_c(j+1) \cdot p[i, j+1] \\
        & - \mu_{\text{TLOF}} \cdot p[i, j] = 0; \\
        & \quad \text{for } i=K, j=0
    \end{aligned} &
\end{flalign}

\begin{flalign}  \label{4th}
    & \begin{aligned}
        & \mu_{\text{TLOF}} \cdot p[i+1, j-1] - (\lambda + \mu_c(j)) \cdot p[i, j] = 0; \\
        & \quad \text{for } i=0, j=c
    \end{aligned} &
\end{flalign}

\begin{flalign}  \label{5th}
    & \begin{aligned}
        & \lambda \cdot p[i-1, j] - \mu_c(j) \cdot p[i, j] = 0; \\
        & \quad \text{for } i=K, j=c
    \end{aligned} &
\end{flalign}

\begin{flalign}  \label{6th}
    & \begin{aligned}
        & \lambda \cdot p[i-1, j] + \mu_c(j+1) \cdot p[i, j+1] \\
        & - (\lambda + \mu_{\text{TLOF}}) \cdot p[i, j] = 0; \\
        & \quad \text{for } 0<i<K, j=0
    \end{aligned} &
\end{flalign}

\begin{flalign}  \label{7th}
    & \begin{aligned}
        & \lambda \cdot p[i-1, j] + \mu_c(j+1) \cdot p[i, j+1] \\
        & - (\mu_{\text{TLOF}} + \mu_c(j)) \cdot p[i, j] = 0; \\
        & \quad \text{for } i=K, 0<j<c
    \end{aligned} &
\end{flalign}

\begin{flalign}  \label{8th}
    & \begin{aligned}
        & \lambda \cdot p[i-1, j] + \mu_{\text{TLOF}} \cdot p[i+1, j-1] \\
        & - ( \mu_c(j) + \lambda) \cdot p[i, j] = 0; \\
        & \quad \text{for } 0<i<K, j=c
    \end{aligned} &
\end{flalign}

\begin{flalign}  \label{9th}
    & \begin{aligned}
        & \mu_c(j+1) \cdot p[i, j+1] + \mu_{\text{TLOF}} \cdot p[i+1, j-1] \\
        & - (\lambda + \mu_c(j)) \cdot p[i, j] = 0; \\
        & \quad \text{for } i=0, 0<j<c
    \end{aligned} &
\end{flalign}

\begin{flalign}  \label{10th}
    & \begin{aligned}
        & \lambda \cdot p[i-1, j] + \mu_{\text{TLOF}} \cdot p[i+1, j-1] \\
        & + \mu_c(j+1) \cdot p[i, j+1] \\
        &- (\mu_{\text{TLOF}} + \lambda + \mu_c(j)) \cdot p[i, j] = 0; \\
        & \quad \text{for } 1<i<K, 1<j<c
    \end{aligned} &
\end{flalign}

The global balance equations capture the steady-state behavior of the system. The first eight equations represent edge cases. These cases represent the boundaries of the model where the behavior is dictated by the constraints of the vertiport's capacity in both the terminal airspace and parking pads. The last equation encapsulates the main dynamics. 

Eqn. \ref{2nd} corresponds to the system state at the upper left corner, the initial state $(0, 0)$. Here, the departure rate from the parking pads ($\mu_c(j+1)$) must be balanced by the aircraft arrival rate to the buffer ($\lambda$), ensuring that the outflow and inflow are in equilibrium when no aircraft are present.

Eqn \ref{3rd} addresses the case for $i=K$ and $j=0$, the bottom-left edge of the system. This represents a full buffer where an incoming aircraft will be rejected to join the queue.

Eqn. \ref{4th} for the state where $i=0$ and $j=c$ describes a situation with an empty buffer and full parking pads. In this case, the TLOF is blocked for service.

Eqn. \ref{5th} is for the state $i=K$ and $j=c$, which is the bottom-right corner of the diagram. This state signifies both full terminal airspace and fully occupied parking pads, limiting the system's ability to accept new arrivals until a departure occurs.

Eqn. \ref{6th} governs the dynamics for the states along the vertical axis where $0<i<K$ and $j=0$. In this scenario, the equation factors in the arrival of an aircraft to an empty parking pad and the departure of aircraft following charging and TLOF service. Similarly, equations \ref{7th}, \ref{8th}, and \ref{9th} represent the edge rows and columns of the state transition rate diagram. These encompass scenarios where aircraft arrivals are confined by the availability of airspace and parking pads.

Finally, the last equation captures the main dynamics of the system for states away from the edges, where $1<i<K$ and $1<j<c$. This equation embodies the complex interplay of aircraft arrival, departure, and the TLOF service rates.

\subsection{Performance Metrics}
Once the global balance equations are solved and the steady-state probabilities $\pi(i, j)$ are obtained, key performance metrics of the vertiport can be evaluated. We are particularly interested in the average queue length in the airspace, the variance of the queue length, the average waiting time in the queue, and the variance of this waiting time.

The average queue length in the airspace, denoted as $L_q$, is a measure of the expected number of aircraft in the terminal airspace queue and is calculated as follows:

\begin{equation}
L_q = \sum_{(i, j) \in E} i \cdot \pi(i, j)
\end{equation}

The variance of the queue length, denoted as $\sigma_q^2$, provides a measure of the dispersion of the queue length from the average. It is computed by:

\begin{equation}
\sigma_q^2 = \sum_{(i, j) \in E} (i - L_q)^2 \cdot \pi(i, j)
\end{equation}

The variance in queue length is important because it quantifies the fluctuation in the number of aircraft waiting over time, which is essential for assessing the reliability of vertiport traffic management. A lower variance indicates a more consistent and predictable queue length which facilitates smoother vertiport operations and planning.

% In the system under study, while  $L_q$ is important metric for maintaining the safety of terminal airspace, we place greater emphasis on the queue waiting time, represented by $W_q$. This preference is due to scenarios where high charging times and limited parking pad availability can lead to considerable waiting times, despite low queue lengths, particularly at high arrival rates. Therefore, $L_q$ is a potentially biased metric whereas $W_q$ serves as a more accurate measure of the actual delays encountered. $W_q$ can be derived from Little's Law:

In the system under study, we take the queue waiting time, represented by $W_q$, as the main metric to assess the performance of the system. $W_q$ can be derived from Little's Law:

\begin{equation}
W_q = \frac{L_q}{\lambda}
\end{equation}

where $\lambda$ is the average arrival rate of aircraft at the vertiport.

The variance in waiting time, denoted as $\sigma_{W_q}^2$, is crucial for assessing the consistency of waiting times, which affects the predictability of vertiport operations. It is given by:

\begin{equation}
\sigma_{W_q}^2 = \frac{\sigma_q^2}{\lambda^2}
\end{equation}

\subsection{Interaction Between Queues}
The exploration of interactions between queues in a vertiport setting reveals intricate dynamics that significantly influence the system's operational efficiency.

The vertiport queueing network model depicted in Fig. \ref{fig:feedback_model} contrasts with a conventional M/M/c queue model by accounting for the limited availability of parking pads, which introduces a blocking mechanism not present in the simpler model. In the traditional M/M/c queuing model, the system assumes an infinite buffer, which fails to capture the practical limitations of vertiport operations.

% Fig. \ref{fig:blocking_probability} illustrates the increase in blocking probability with higher arrival rates ($\lambda$) for various numbers of parking pads, given a fixed charge rate ($\mu_{charge}$) of 6 aircraft/hour. The solid lines represent scenarios where the average queue length ($L_q$) remains at or below 3. Blocking can occur from two queues within the system. The first instance of blocking can occur in the terminal airspace when the buffer is full (in our case, with $K=100$), and the second blocking can occur when parking pads are full. Fig. \ref{fig:blocking_probability} represents the blocking probability that occurs from the second queue. Consequently, the probabilities do not sum to 1, as the first queue's blocking probability is not included. Incorporating the blocking probability from the first queue would complete the total to 1. Since the buffer size of the second queue is considerably lower than that of the first queue, blocking initially occurs in the second queue. As we aim to demonstrate that isolating two queues does not accurately reflect reality, we focus solely on the blocking probability of the second queue. 

\begin{figure}[ht]
\centering
\includegraphics[width=0.45\textwidth]{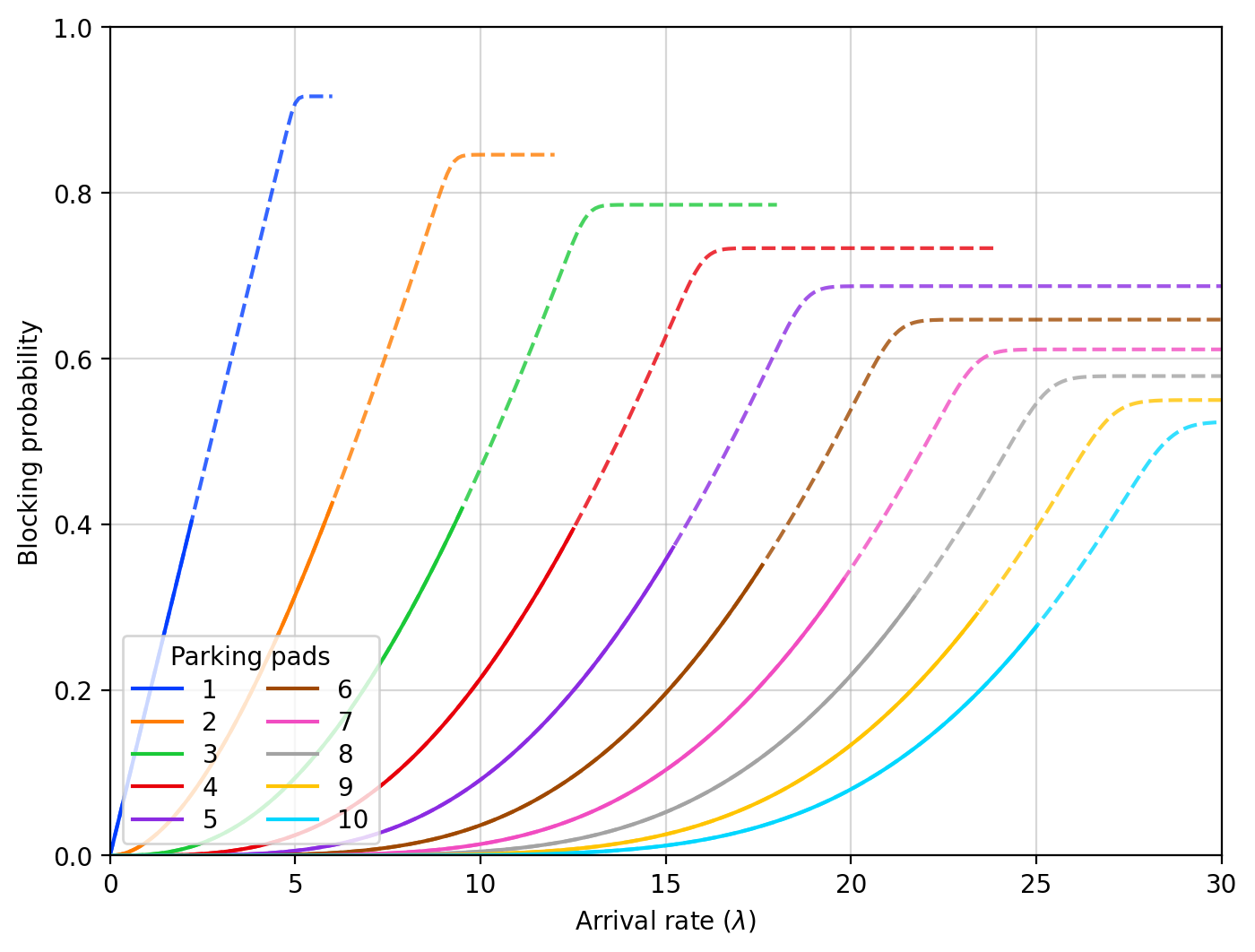}
\caption{Blocking probability of the parking pads when $\mu_{charge} = 6$ aircraft/hour. Solid lines where $W_q \leq 10$ minutes}
\label{fig:blocking_probability}
\end{figure}

Fig. \ref{fig:blocking_probability} illustrates how the blocking probability increases with higher arrival rates ($\lambda$) for different numbers of parking pads, under a constant charge rate ($\mu_{charge}$) of 6 aircraft per hour. The solid lines represent scenarios where the average queueing delay ($W_q$) remains at or below 20 minutes at the first queue (TLOF arrival queue). Given that the capacity of the terminal airspace is significantly high ($K$=100), blocking in this first queue is relatively negligible. Conversely, the capacity of the parking pads is an order of magnitude lower, making it the primary source of blocking within the system. Therefore, Fig. \ref{fig:blocking_probability} specifically presents the blocking probability associated with the second queue, where the parking pads are located. As a result, the probabilities shown do not sum to 1 because the blocking probability from the first queue is excluded. If it were included, the total blocking probabilities from both queues would sum to 1.

% Notably, it is observed that the blocking probability surpasses 20\% (30\%) when the arrival rate approximates 40\% (50\%) of the designated surface capacity ($\mu_{charge} \cdot c$) for a majority of the parking pad configurations. Furthermore, when the steady-state average queueing delay remains at or below 20 minutes, the blocking probability hovers at approximately 40\% or higher, even in scenarios with up to ten parking pads. This probability escalates substantially as the number of parking pads decreases, peaking at 60\% for a vertiport with a single parking pad configuration. This means that when the vertiport is operated at $\sim$50\% of the surface capacity, 

The analysis reveals that even with a steady-state average queueing delay kept around 10 minutes, the blocking probability is 30\% or higher for all parking pad configurations. This probability escalates as the number of parking pads decreases, peaking at over 40\% for a vertiport with a single parking pad configuration. Notably, when the blocking probability exceeds 30\%, the arrival rate only reaches about 50\% of the designated surface capacity ($\mu_{charge} \cdot c$) for most configurations. These observations imparts two critical insights. Firstly, they highlight the operational challenges in vertiport design; specifically, the necessity to keep the arrival rate significantly below the surface capacity to manage the incidence of blocking events effectively. The observed high blocking probabilities indicate substantial interaction between queues: even with low queueing delays, the second queue significantly impacts the first queue's performance, affecting it 40\% of the time under the studied conditions. Secondly, the assumptions of Jackson networks, such as Poisson arrivals and exponential service times that allow for the isolation and independent analysis of each queue \cite{jackson1963jobshop}, are not applicable for our system. Due to the blocking of the second queue, the arrivals to the second queue do not form a Poisson process. Therefore, our analysis underscores the need for a holistic analysis of the queueing network structure, as the secondary queue contributes to blocking in the primary queue considerably before the arrival rate reaches the full surface capacity.

\begin{figure}[ht]
\centering
\includegraphics[width=0.45\textwidth]{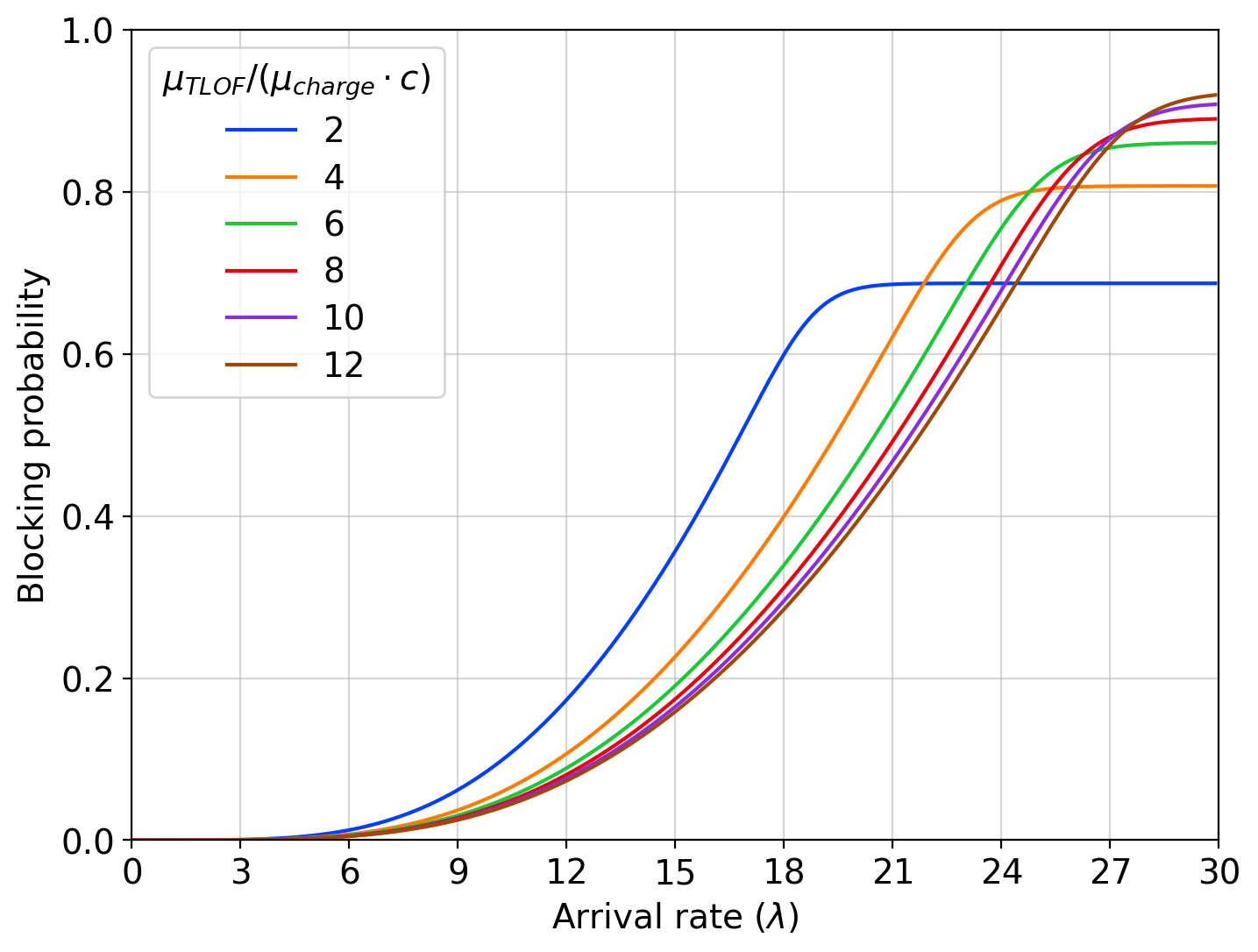}
\caption{Blocking probability of the parking pads when $\mu_{charge} = 6$ aircraft/hour and $c = 5$.}
\label{fig:blocking_prob_for_tlof_park_ratio}
\end{figure}

Additionally, Fig. \ref{fig:blocking_prob_for_tlof_park_ratio} maintains the product of the charge rate and the number of parking pads at a constant value of 30, and it shows how variations in the TLOF service rate ($\mu_{TLOF}$) affect the blocking probability. A noteworthy observation is that the change in the blocking probability becomes marginal when the ratio $\mu_{TLOF}/(\mu_{charge} \cdot c)$ exceeds 8, implying that the interaction between the TLOF queue and the parking pad queue diminishes. At this point, it is conceivable to consider the queues as being sufficiently independent to decompose both queues and apply simple M/M/1 and M/M/c models for performance analysis. However, such independence is unachievable with the operational parameters proposed in our work and in existing literature, as the required ratio of 8 is not attainable \cite{vascik2019development, wisk_conops}. These necessitate solving the system as a whole and consider the interaction between the queues.

\section{Simulation Model} \label{sim_model}
For this study, we utilized VertiSim to simulate vertiport operations \cite{scitech_vertisim, onat2024simulationoptimization}. VertiSim leverages SimPy, a discrete-event simulation framework in Python \cite{SimPy}. SimPy's event-driven approach enables the representation of the vertiport's operations, including aircraft arrivals, landings, charging completions, departures, and TLOF availability changes. Events advance the simulation clock and trigger state changes within the system. In the simulation, arriving aircraft join the TLOF arrival queue. An aircraft can only proceed when it secures a parking pad and reserves the TLOF. The simulation implements the BBS-SNO mechanism, where aircraft must wait in the arrival queue if no parking pads are available, until a space is freed by another aircraft departing via the TLOF. After servicing at the TLOF and completing charging at a parking pad, aircraft then queue for TLOF access for departure. This introduces a departure queue in the simulation model, contrasting with the queueing model where aircraft directly transition from parking pad service to TLOF service.

\subsection{Simulation Validation}
We validated our simulator using the queueing model illustrated in Fig. \ref{fig:simplified_model}. For simulation and queueing models, we used the parameters of $\mu_{\text{TLOF}} = 30$ aircraft/hr, $\mu_{\text{charge}} = 6$ aircraft/hr, and the number of parking pads ranged from 1 to 10. We compared queuing delay metrics between the simulated and theoretical models. As shown in Table \ref{tab:sim_validation}, the results confirm that the simulation model closely matches the mathematical results, with the largest observed difference between the queueing delays being a mere 5.9 seconds (3\% maximum difference). 

Given helicopter regulations, VFR conditions mandate enough fuel reserves to permit an additional 20 minutes of cruising beyond the planned mission \cite{faa_vfr_rule}. It is anticipated that eVTOL aircraft will have similar reserve energy requirements \cite{HATHERALL2024122848}. However, considering the specific constraints of UAM, where the proposed flight times are relatively short and batteries are limited \cite{scitech_vertisim, archerArcherAnnounces}, it is impractical to accommodate a queueing delay close to these reserve limits. In our scenarios, even if the queueing delay approached 10 minutes, our simulation's approximation error would translate to a maximum of only 18 seconds. With this high level of accuracy, we can confidently employ the simulator to model more complex scenarios and non-Markovian distributions.

\begin{table}[ht]
\begin{center}
\begin{tabular}{
  |c?{0.5mm}c|c|c| % Four initial columns
  l % The problematic 'l' turned into a normal column without vertical line before the thicker line
  ?{0.5mm} % Specifies the thick vertical line
  c|c|c|l| % Four columns after the thick line
}
\hline
$\lambda$ & c & $W_q^a$ & $W_q^s$ & \multicolumn{1}{c?{0.5mm}}{Diff} & c  & $W_q^a$ & $W_q^s$ & \multicolumn{1}{c|}{Diff} \\ \specialrule{1.5pt}{0.5pt}{0.5pt}
1         & 5 & 0.069   & 0.070   & 0.001                           & 10 & 0.069   & 0.070   & 0.001                     \\ \hline
2         & 5 & 0.143   & 0.143   & 0.000                           & 10 & 0.143   & 0.142   & 0.001                     \\ \hline
3         & 5 & 0.223   & 0.223   & 0.000                           & 10 & 0.222   & 0.222   & 0.000                     \\ \hline
4         & 5 & 0.309   & 0.312   & 0.003                           & 10 & 0.308   & 0.307   & 0.001                     \\ \hline
5         & 5 & 0.406   & 0.416   & 0.010                           & 10 & 0.400   & 0.401   & 0.001                     \\ \hline
6         & 5 & 0.519   & 0.522   & 0.003                           & 10 & 0.500   & 0.501   & 0.001                     \\ \hline
7         & 5 & 0.658   & 0.664   & 0.006                           & 10 & 0.609   & 0.609   & 0.000                     \\ \hline
8         & 5 & 0.841   & 0.904   & 0.063                           & 10 & 0.727   & 0.727   & 0.000                     \\ \hline
9         & 5 & 1.100   & 1.125   & 0.025                           & 10 & 0.857   & 0.856   & 0.001                     \\ \hline
10        & 5 & 1.490   & 1.514   & 0.024                           & 10 & 1.001   & 0.999   & 0.002                     \\ \hline
11        & 5 & 2.111   & 2.137   & 0.026                           & 10 & 1.160   & 1.160   & 0.000                     \\ \hline
12        & 5 & 3.151   & 3.249   & 0.098                           & 10 & 1.341   & 1.343   & 0.002                     \\ \hline
\end{tabular}
\end{center}
\caption{Comparison of analytical and simulation model results. $\lambda$ indicates the arrival rate (aircraft/hr), $W_q^a$ represents the queueing delay from the analytical model, and $W_q^s$ represents the queueing delay from the simulation. The 'Diff' column shows the absolute difference between the two models' results, with all times in minutes.}
\label{tab:sim_validation}
\end{table}

\section{Results} \label{results}
This section presents the findings from the queuing model and simulation model used to study vertiport operations. In the queuing analyses, which employ a two-node queuing network model, we held $\mu_{TLOF}$ constant at 30 operations per hour. For the simulation analyses, which model the feedback system, we held $\mu_{TLOF}$ at 60 operations per hour. As detailed in section \ref{vertiport_ops_model}, both of these rates correspond a total of 60 arrivals and departures to the TLOF per hour. Once the TLOF capacity is set, other design parameters like the charge rate and number of parking pads can be selected according to specific operational needs. For example, a vertiport designer could set the number of parking pads and determine the necessary charge rate to achieve a desired throughput level based on the mean and variance of the required charging times, or conversely, set the total power draw from the electrical grid and determine the number of parking pads required to support it for their target throughput level.

In our analysis, we determine the maximum achievable throughput ($\Lambda$) while ensuring that $W_q$ remains below a designated threshold with a specified confidence level. This will ensure, for example, that the aircraft waiting time will be less than 5 minutes with 95\% probability. The analytical distribution of waiting times is difficult to obtain however \cite{balsamoBook}. Therefore, instead of utilising the analytical distribution, we model $W_q$ as a normally distributed random variable. This assumption is justified by similar approximations from \cite{Gomez2004} and \cite{Grottke2011} where waiting times of tandem queues were approximated using normal distributions. This assumption will be further justified in Table \ref{tab:throughput_validation} where analytical and simulation results concord. 

To achieve this, we determine mean and variance of the waiting time $W_q$ using our analysis and Little's law. The the upper confidence limit of $W_q$ is then obtained using the following equation:

\begin{equation} \label{eqn:confidence}
    W_q^{upper} = \overline{W_q} + Z \times \sigma_{Wq}.
\end{equation}

Here Z is the Z-score corresponding to the desired confidence level obtained from the standard normal distribution \cite{Devore_2012}. By iterating through various $\lambda$ values, we identify the maximum $\lambda$ for which $W_q^{upper}$ does not exceed the threshold, thus providing a statistically robust measure of system performance. In contrast, for simulations, we can directly obtain the necessary percentile using the data without computing the upper confidence limit. For all analyses that require specifying a confidence level, we set it at 95\% for the maximum queuing delay of 5 minutes.

\subsection{Queueing Model Results}
\begin{figure}[ht]
\centering
\includegraphics[width=0.45\textwidth]{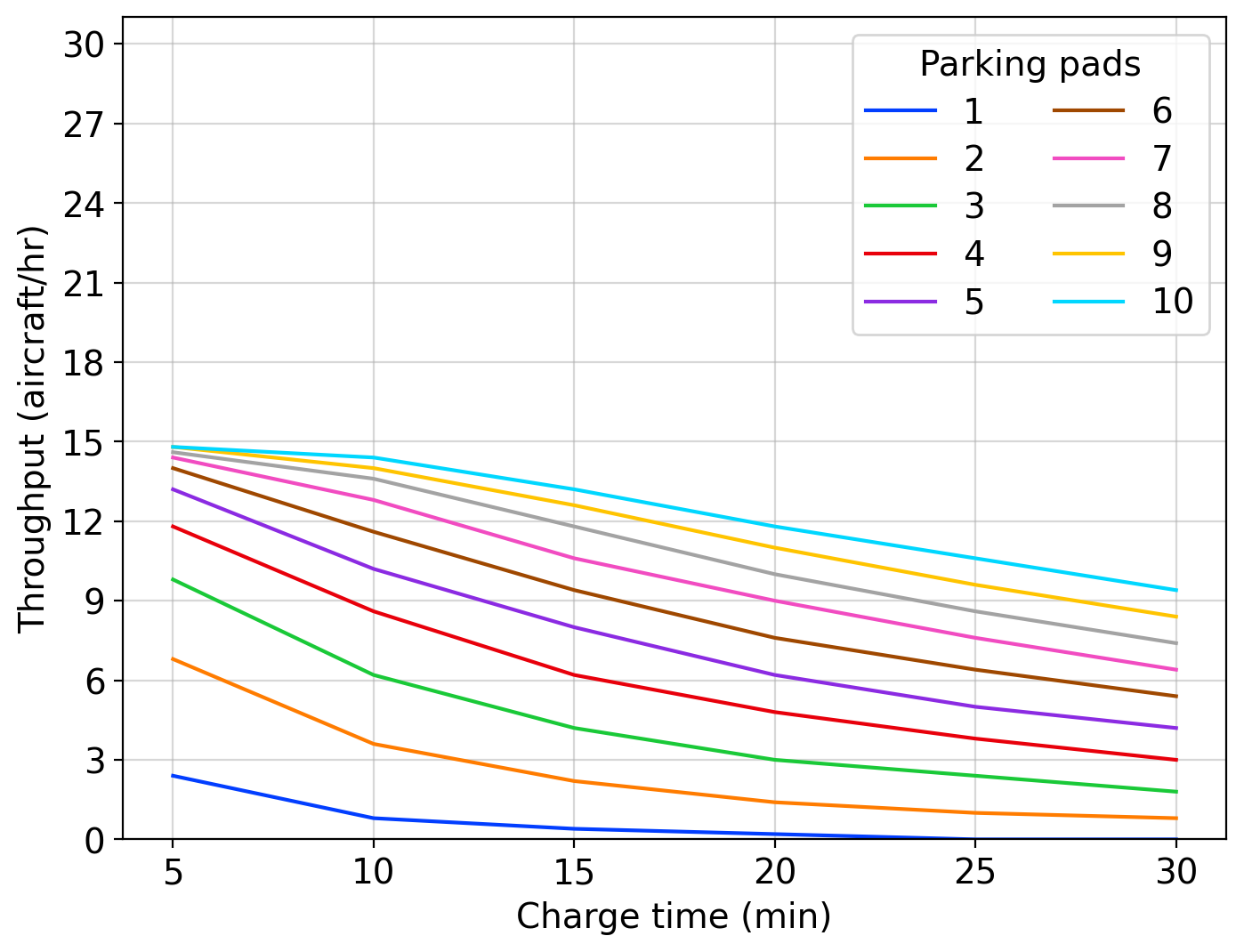}
\caption{Maximum achievable throughput for parking pads across charge times from 5 to 30 minutes, with a 95\% confidence level and a maximum $W_q$ of 5 minutes}
\label{fig:queuein_analysis_throughput}
\end{figure}

Figure \ref{fig:queuein_analysis_throughput} illustrates the impact of varying the number of parking pads and charge times on the maximum achievable throughput at a vertiport, ensuring that $W_q$ remains below 5 minutes with 95\% confidence. Each line represents a different number of parking pads. We observe an exponential relationship between charge time and throughput for all the number of parking pads. The plot reveals that at lower charge times, increasing the number of parking pads leads to a higher increase in throughput. This is because shorter charge times allow for quicker turnaround and enable the system to handle a higher volume of aircraft. However, as charge times extend, the incremental throughput gain per added parking pad diminishes. For instance, with a 30-minute charge time, the throughput difference between the minimum and maximum number of parking pads is an increase of 9 aircraft per hour. In contrast, with a 5-minute charge time, this difference expands to 13 aircraft per hour.

Additionally, with shorter charging times and an increased number of parking pads, the system quickly approaches the total TLOF capacity, which demonstrates diminishing returns more swiftly than at longer charging times. We observe that, except for a 5-minute charge time, throughput increases linearly with the number of parking pads. Beyond this point, the relationship between throughput and the number of parking pads becomes nonlinear for other charge durations as throughput nears the TLOF's arrival or departure capacity. Notably, after the addition of seven parking pads, any further increase in throughput becomes negligible. Ultimately, the maximum capacity of the vertiport is found to be either half the service rate of the TLOF or slightly less than 20\% of the surface capacity when using the maximum effective number of parking pads (7).

We remark that the 

To determine the accuracy of the two-node queueing model when predicting the behavior of the feedback queueing model, we use our simulator. We conducted a series of simulations with 200 distinct seed values to ensure a robust stochastic representation. For simplicity, we present results for configurations with 5 and 10 parking pads in Table \ref{tab:sim_validation}. The analytical results closely follow the simulation curves, with a maximum deviation of 10\%. These results indicate that the two-node approximation is good representation of the feedback system and would be able to inform vertiport design.

\begin{table}[ht]
\begin{center}
% \label{tab:sim_validation}
\begin{tabular}{|c?{0.5mm}c|c|c|c?{0.5mm}c|c|c|c|}
\hline
$T_{charge}$ & c & $\Lambda^a$ & $\Lambda^s$ & Diff & c  & $\Lambda^a$ & $\Lambda^s$ & Diff \\ \specialrule{1.5pt}{0.5pt}{0.5pt}
5            & 5 & 13.2        & 14          & 0.8  & 10 & 14.8        & 15.6        & 0.6  \\ \hline
10           & 5 & 10.2        & 10.6        & 0.4  & 10 & 14.4        & 15.1        & 0.7  \\ \hline
15           & 5 & 8.0         & 7.7         & 0.3  & 10 & 13.2        & 14.6        & 1.4  \\ \hline
20           & 5 & 6.2         & 6.1         & 0.1  & 10 & 11.8        & 13.2        & 1.4  \\ \hline
25           & 5 & 5.0         & 5.1         & 0.1  & 10 & 10.6        & 11.7        & 1.1  \\ \hline
30           & 5 & 4.2         & 3.9         & 0.3  & 10 & 9.4         & 10.1        & 0.7  \\ \hline
\end{tabular}
\end{center}
\caption{Comparison of analytical and simulation model results for maximum achievable throughput calculation ($\Lambda$). The superscripts 'a' and 's' denote results from the analytical and simulation models, respectively. $T_{charge}$ denotes charging times in minutes, and 'c' represents the number of parking pads in the system.}
\label{tab:throughput_validation}
\end{table}

\begin{figure}[ht]
\centering
\includegraphics[width=0.45\textwidth]{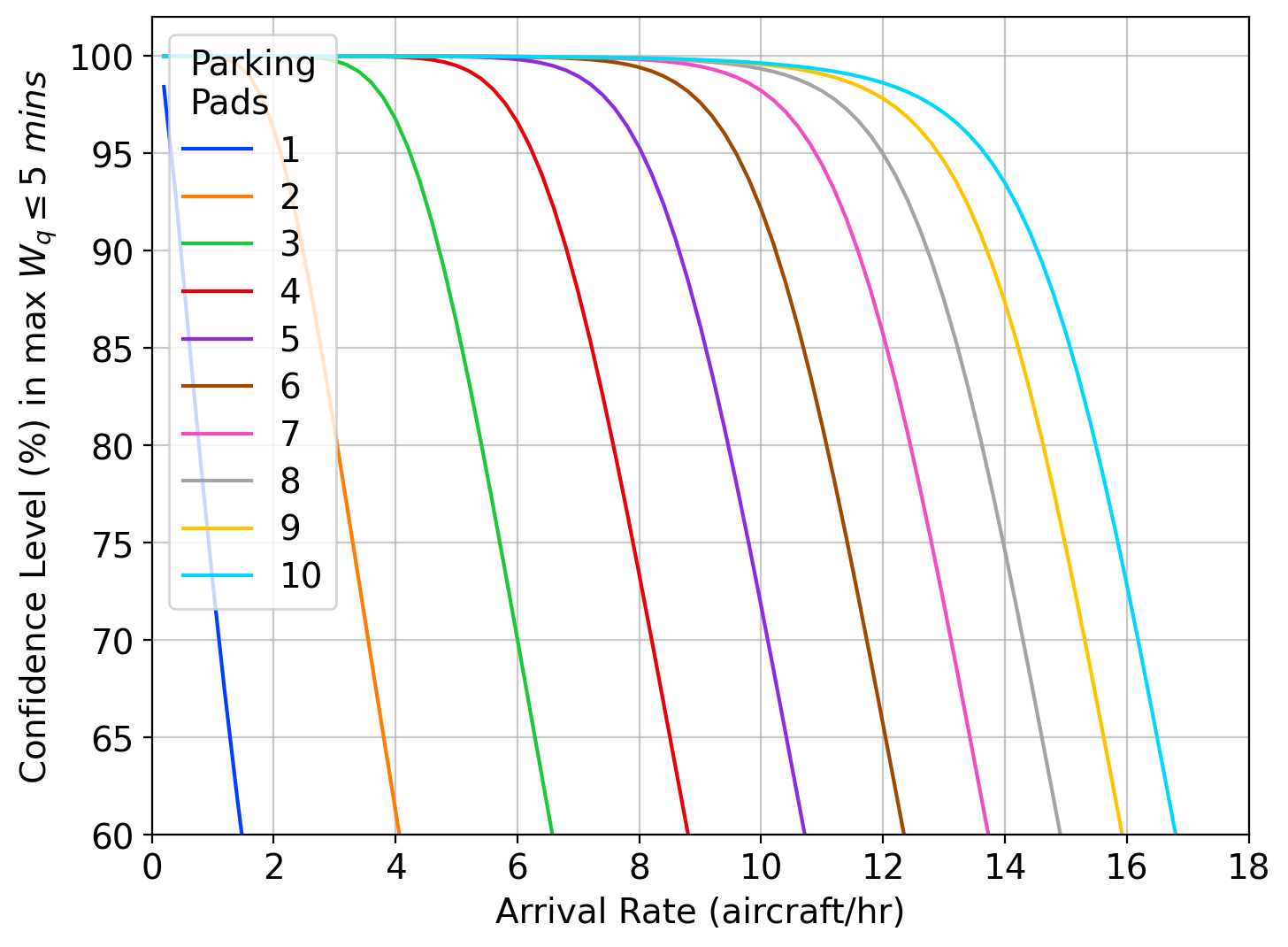}
\caption{Confidence levels in maintaining a maximum queue waiting time of 5 minutes at different arrival rates for 15 minutes of charging time}
\label{fig:confidence_levels}
\end{figure}

Fig. \ref{fig:confidence_levels} illustrates the confidence levels associated with maintaining $W_q$ of 5 minutes at varying arrival rates for configurations ranging from 1 to 10 parking pads. Each curve represents a different number of parking pads and shows the confidence level in percent as a function of the arrival rate (aircraft/hour). The plot highlights that for each configuration, as the arrival rate increases, the system initially maintains a high confidence level. However, once the arrival rate surpasses a certain threshold, which is specific to each parking pad configuration, the confidence level declines sharply. This sharp decline indicates a significant reduction in the predictability of the system's performance. Beyond this threshold, further increases in the arrival rate do not substantially enhance system throughput; rather, they diminish the system's controllability.

Given the findings, vertiport designers should be aware that controlling the system effectively becomes challenging beyond certain arrival rates. It is crucial to design vertiports with these considerations in mind and to ensure that the arrival rates do not exceed the threshold where the system's predictability significantly decreases.

% Given the maximum ratio of variance to average queue length set at 4, Fig. \ref{fig:charge_c_throughput} reveals that at the configurations with lower number of parking pads, the throughput is around at the 50\% of the surface capacity and independent of the charge rate. In contrast, at higher parking pad configurations, the throughput remains around 50\% of the surface capacity when charge rates are low. Yet, the throughput dips below 25\% as the charge rates increase. These findings quantify a notable inefficiency that arises with an increased number of parking pads at higher charge rates. 

\begin{figure}[ht]
\centering
\includegraphics[width=0.45\textwidth]{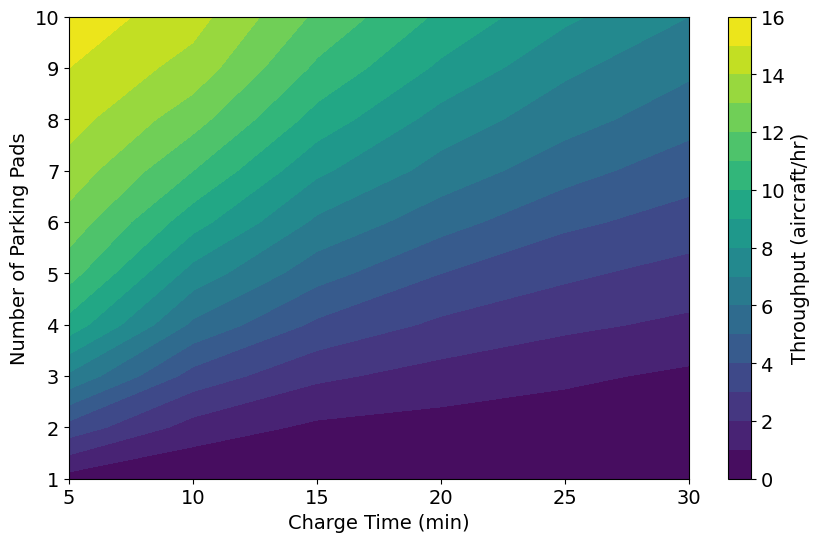}
\caption{Feasible throughput regions under a constraint of keeping max $W_q$ below 5 minutes with 95\% confidence. Throughput varies from low (dark blue) to high (yellow).}
\label{fig:area_plot}
\end{figure}

Fig. \ref{fig:area_plot} highlights the trade-off between achieving high throughput and maintaining design flexibility at a vertiport. This plot illustrates the feasible throughput regions of a vertiport as a function of both the charge time and the number of parking pads, under the confidence constraint. Notably, moving left along the x-axis (reducing charge time) and up the y-axis (increasing the number of parking pads) leads to higher throughput levels. However, this region of higher throughput is increasingly restricted in terms of feasible design options. In contrast, moving right along the x-axis (increasing charge time) and down the y-axis (reducing the number of parking pads) offers more flexibility and a broader range of feasible options, though at lower throughput levels.

\subsection{Non-Markovian Model Results}

%Mention that runway service does not have such high variance. So, it is acceptable to assume a deterministic number, poisson arrivals are already validated. Justify the normal distribution for charging times.

%Simulation allows us to study non-exponential distributions and more complex systems that queueing theory does not allows us to model.

% Although aircraft arrivals following a Poisson process have been validated in airport settings~\cite{willemain2004}, most vertiport operations do not have exponentially-distributed times. Indeed, runway service times, for example, have a low enough variance that it is acceptable to assume a deterministic time \cite{}. Likewise, the charging time of aircraft in a two-vertiport system should follow a normal distribution rather than an exponential distribution. This can be explained by the set distance to be covered with random variations such as wind or delays.

% WE SHOULD NAME THE NORMALLY DIST. CONFig. THEN REFER HERE. OTHERWISE, IT IS HARD TO WRITE

% Results
% 1. Non-markovian --> less variance --> more capacity (Fig. 12 but mention half of it)
% 2. 2 TLOFs --> More throughput with config 2 but they both reach to full capacity if there are enough parking pads
% 3. Config 1 beats config 2 in terms of unit throughput because of the footprint requirement.

Queueing theory often assumes exponentially distributed interarrival and service times, as this leads to tractable analytical solutions. While some models (e.g., M/G/1) allow for closed-form solutions with non-exponential service times, more complex queueing networks usually do not have such solutions. In particular, queueing networks with blocking further complicate the analysis, as they introduce state dependencies and constraints that typically require simulation or approximation techniques for performance evaluation. In airport literature, it has been validated that aircraft arrivals follow Poisson distribution \cite{willemain2004} but not for departures \cite{simaiakis2016queuing}. Therefore, in this section, we use our validated simulator to examine queueing results for more realistic, non-Markovian distributions.

\begin{figure}[ht]
\centering
\begin{minipage}{.23\textwidth}
  \centering
  \includegraphics[width=0.6\linewidth]{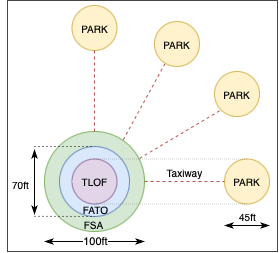}
  \captionof{figure}{Single TLOF layout (Config. 1)}
  \label{fig:single_satellite}
\end{minipage}%
\hspace{0.1cm}
\begin{minipage}{.23\textwidth}
  \centering
  \includegraphics[width=1.05\linewidth]{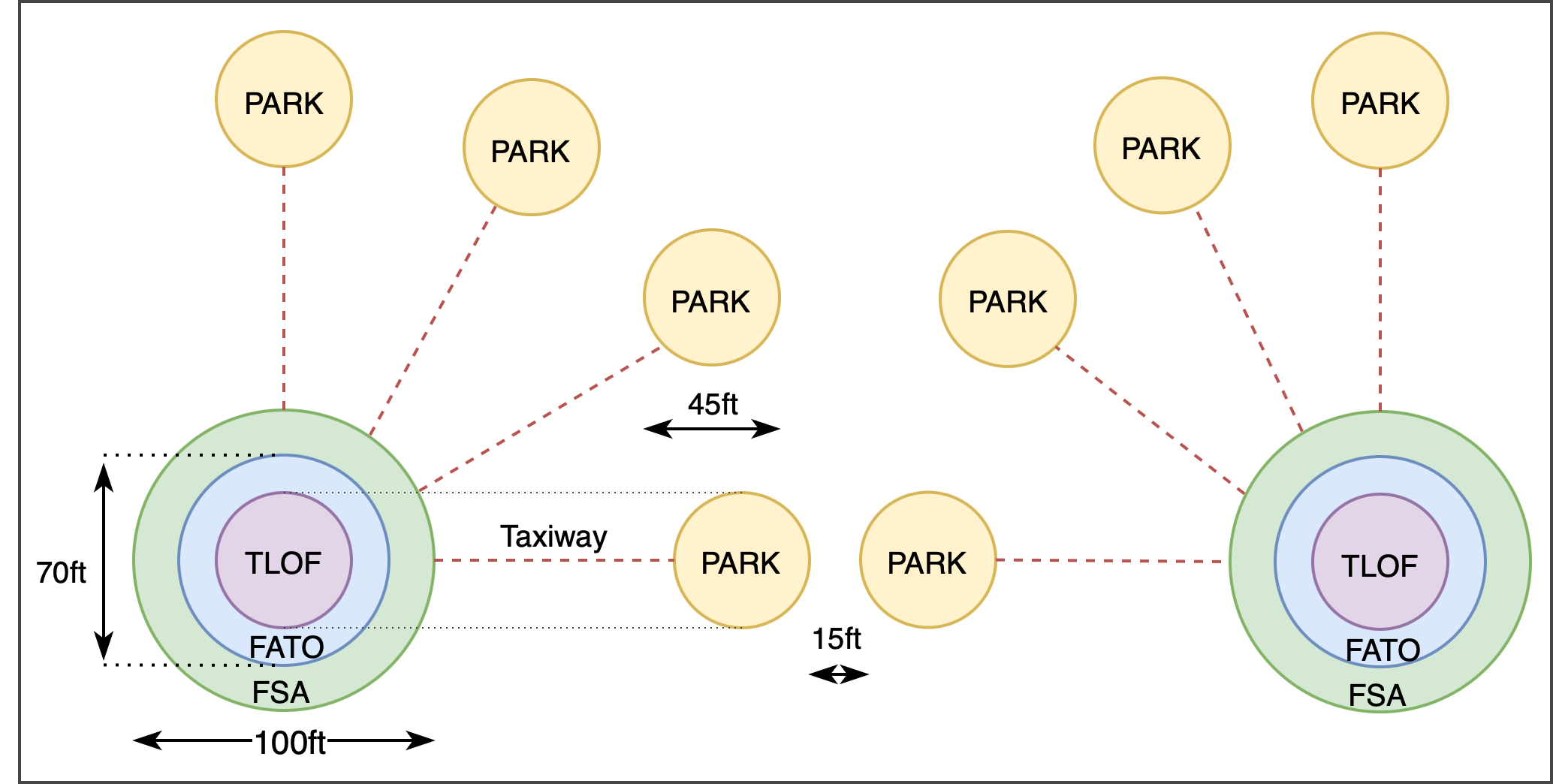}
  \captionof{figure}{Two independent TLOFs (Config. 2)}
  \label{fig:satellite_normal}
\end{minipage}
\end{figure}

\begin{figure}[ht]
\centering
\includegraphics[width=0.35\textwidth]{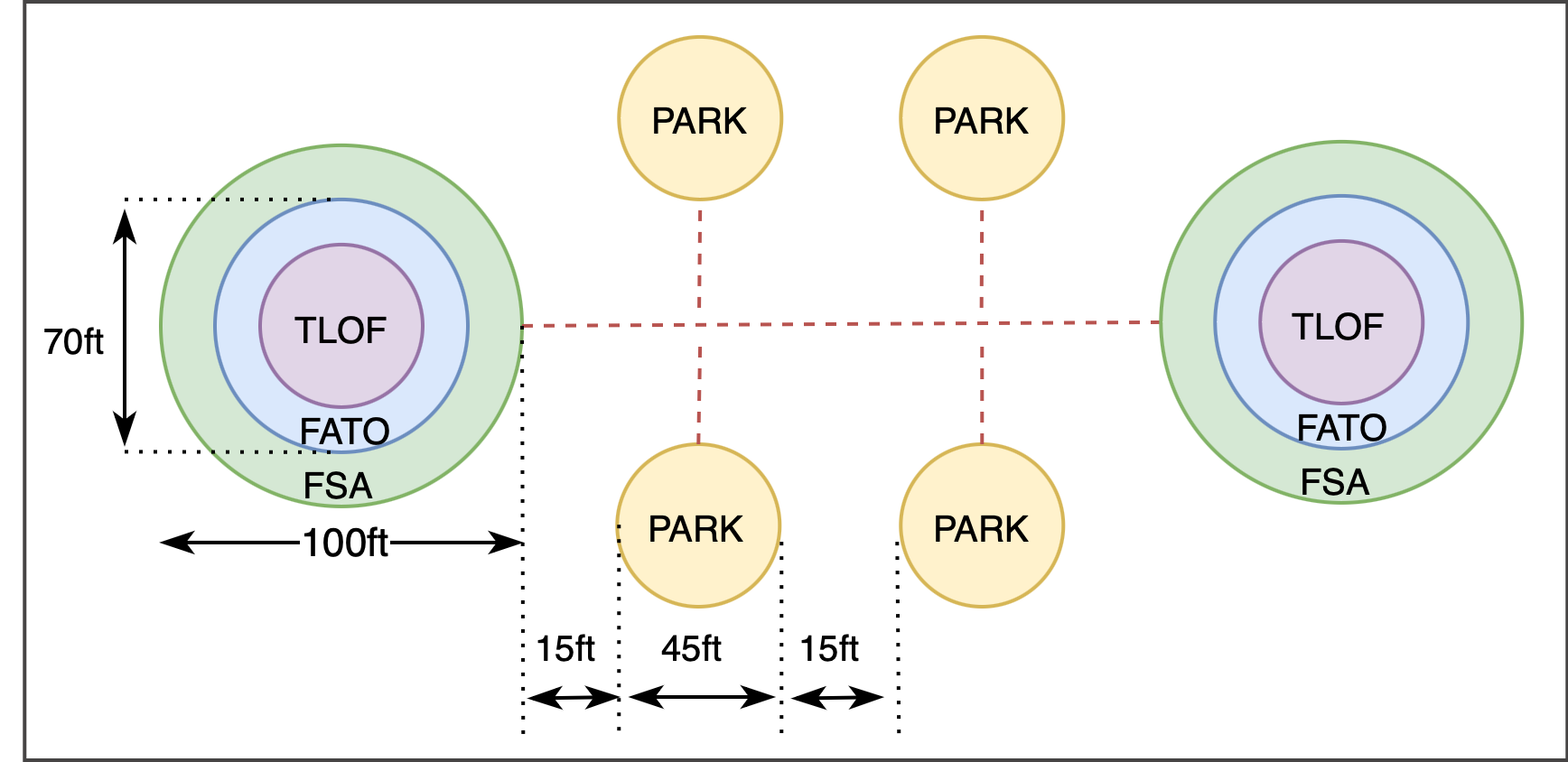}
\caption{Separate TLOFs for takeoff and landing (Config. 3)}
\label{fig:separate_tl}
\end{figure}

In this model, the arrivals follow a Poisson process, TLOF service times are deterministic, and charging times are modeled with a normal distribution, where the standard deviation is set at 20\% of the mean. We considered three distinct vertiport layouts: a single TLOF layout, where one TLOF is surrounded by parking pads in a satellite configuration (Fig. \ref{fig:single_satellite} - Configuration 1); a layout with two independent TLOFs, which doubles the elements of Configuration 1 (Fig. \ref{fig:satellite_normal} - Configuration 2); and a layout featuring separate dedicated takeoff and landing TLOFs (Fig. \ref{fig:separate_tl} - Configuration 3). Config. 3 is also referred as pier layout in the literature \cite{vascik2019development}. We assume that the variance in charging time accounts for the differences in taxi times between parking pads in configuration 3. Aircraft were served using a first-come first-served policy.

% \begin{figure}
% \centering
% \includegraphics[width=0.35\textwidth]{img/satellite_normal.png}
% \caption{Two independent TLOFs (configuration 2)}
% \label{fig:satellite_normal}
% \end{figure}

In the following plots, to be able to make comparison at the same scale with the previous results, we halve the values of two TLOF configurations and present throughput values per TLOF.

\begin{figure}[ht]
\centering
\includegraphics[width=0.45\textwidth]{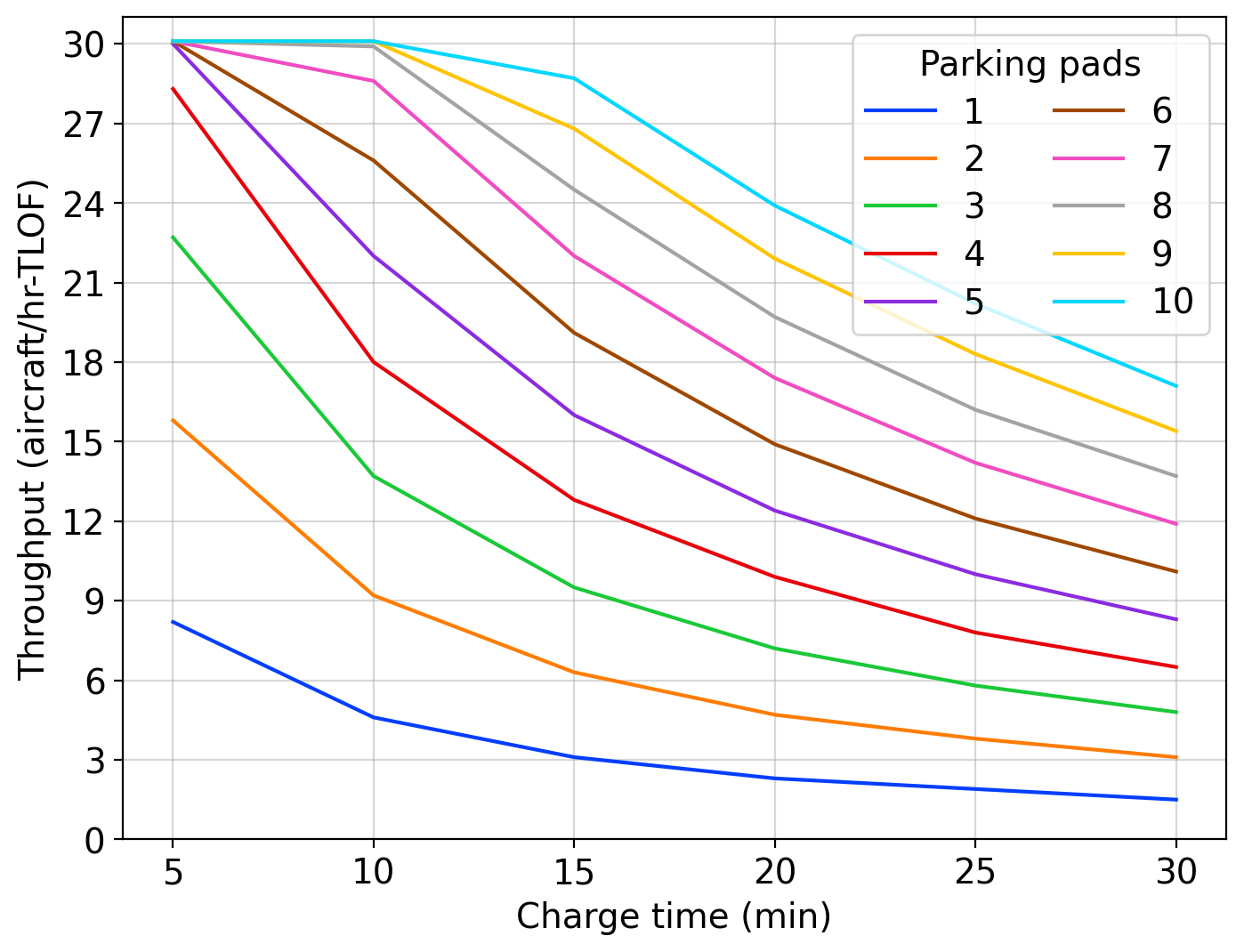}
\caption{Maximum sustainable throughput for Configuration 2 with a 95\% confidence in maximum $W_q <= 5$ minutes}
\label{fig:normal_iso_throughput}
\end{figure}

The result presented in Fig. \ref{fig:normal_iso_throughput} is the maximum throughput obtained from a single TLOF system (Config. 1). Since Config. 2 just doubles the elements for Config. 1, it has the same per TLOF throughput results.
Comparing the values of Fig. \ref{fig:normal_iso_throughput} with Fig. \ref{fig:queuein_analysis_throughput} reveals how dramatic is the cost of the high variance in the operational parameters. The smaller variances when compared to the Markovian case, the TLOF capacity is totally utilized for both layouts for large numbers of parking pads. These simulations should be viewed as the best-case throughput scenario with minimal operational disruption as we keep the variance low at 20\% of the mean charging time.

% The result of the two isolated TLOF vertiport is presented in Fig. \ref{fig:satellite_normal} and the result of the separate TLOF layout is presented in Fig. \ref{fig:separate_tl}. 

\begin{figure}[ht]
\centering
\includegraphics[width=0.45\textwidth]{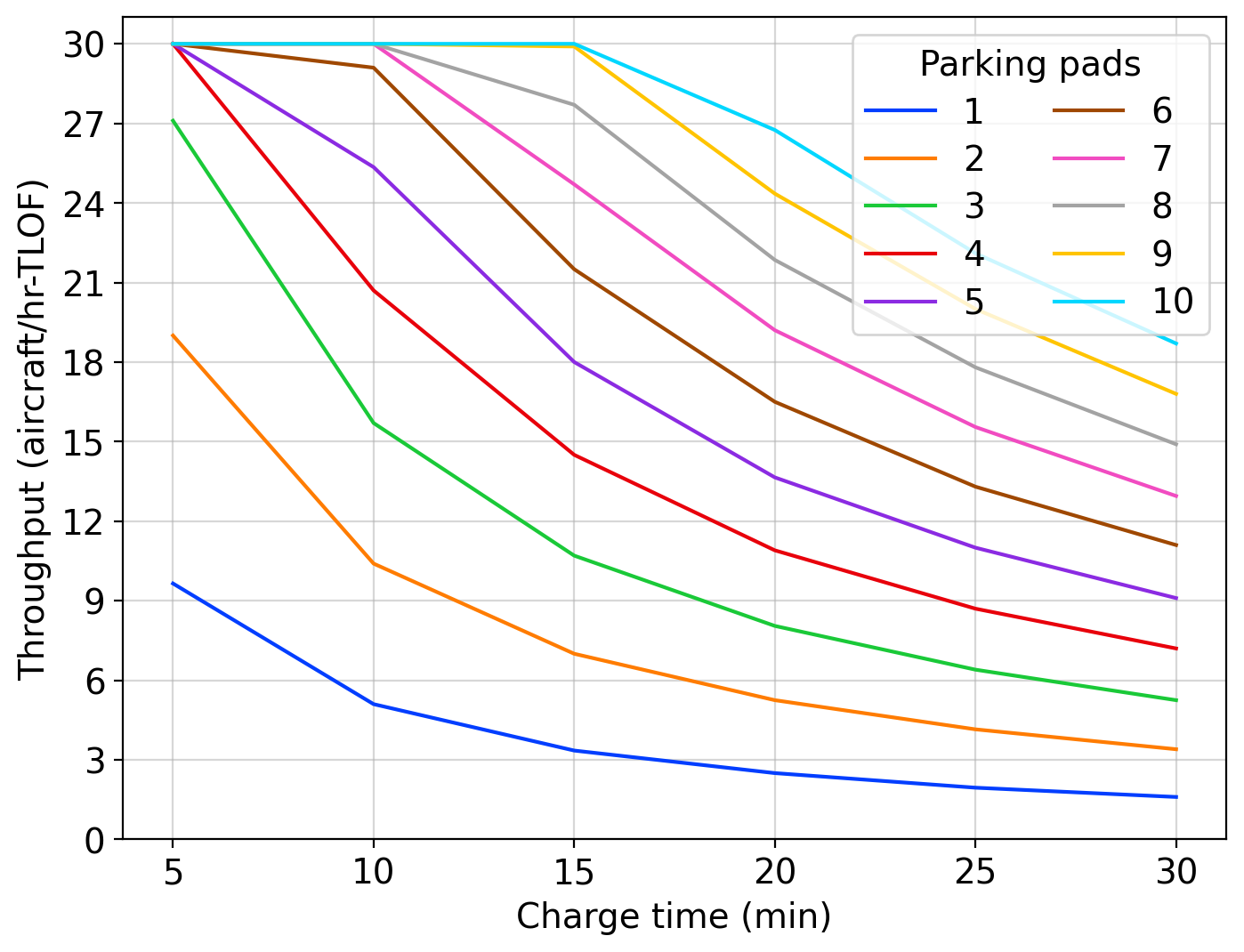}
\caption{Maximum sustainable throughput for Configuration 3 with a 95\% confidence in maximum $W_q <= 5$ minutes}
\label{fig:normal_dedicated_throughput}
\end{figure}

Configuration 3 outperforms Configuration 2, as demonstrated in Figures \ref{fig:normal_iso_throughput} and \ref{fig:normal_dedicated_throughput}. For instance, with six parking pads, Configuration 3 achieves a throughput of 27 aircraft per hour per TLOF, whereas Configuration 2 only manages 23 aircraft per hour per TLOF. Additionally, the maximum effective number of parking pads for Configuration 3 is 5, compared to 4 for Configuration 2 per TLOF. Next, we discover throughput per unit area for each configuration.

% \subsection{Design Suggestions}
\subsubsection{Unit Throughput}
Unit throughput is a key indicator of operational efficiency at vertiports, defined as $U = \frac{\Lambda}{A}$, where $\Lambda$ represents the maximum achievable throughput and $A$ denotes the total area of the vertiport in square feet. This area calculation includes the spacing requirements detailed in \cite{ShannonVertiportTopology} and represents the minimum rectangular area encompassing all vertiport elements. In Configuration 1, the area incrementally increases with each additional parking pad up to nine, as the diameter of the vertiport does not need to expand due to the minimum separation requirements. Beyond this, each new parking pad necessitates a diameter increase to satisfy these spacing criteria. Configuration 2, which doubles the elements of Configuration 1, adheres to guidelines from \cite{heliportdesign} that dictate a minimum separation of 200 feet between FATOs for simultaneous TLOF operations, affecting its spatial layout. Similarly, Configuration 3 maintains 200-foot separations between TLOFs but requires a greater separation distance beyond four parking pads to accommodate additional units, extending only along the horizontal axis.

\begin{figure}[ht]
\centering
\includegraphics[width=0.45\textwidth]{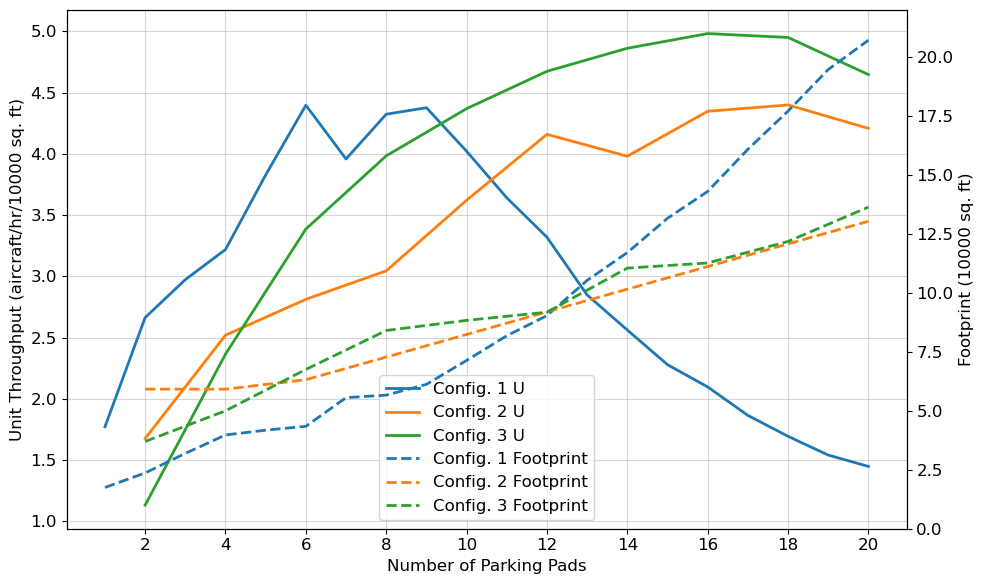}
\caption{Throughput per unit area for Config. 1, 2, and 3}
\label{fig:unit_throughput}
\end{figure}

The unit throughput for each layout configurations are presented in Fig. \ref{fig:unit_throughput}. We observe that Configuration 2 mimics Configuration 1, albeit in a shifted and expanded manner. For Configuration 1, the optimal number of parking pads is six, beyond which the unit throughput stabilizes, but the increasing footprint may diminish its benefits. Configuration 2 remains effective up to 12 parking pads, after which the throughput plateaus. It outperforms Configuration 3 in terms of both throughput and footprint efficiency at or below four parking pads. Overall, Configuration 1 proves to be the most efficient up to nine parking pads but becomes less space-efficient thereafter. The results indicate that Configuration 3 offers the best scalability in terms of throughput per unit area if land availability is not a constraint.

% \begin{figure}
% \centering
% \includegraphics[width=0.45\textwidth]{img/sim_results.png}
% \caption{Queue length in the airspace buffer for $\mu_{charge} = 6$ aircraft/hr as a function of arrival rate and number of parking pads (simulation)}
% \label{fig:sim_result}
% \end{figure}

\section{Conclusion} \label{conclusions}

In this paper, we presented two-node queueing network model with finite buffer and simulation model to determine the maximum sustainable departing aircraft per hour as a function of structural design parameters such as the number of parking pads, number of TLOFs, charger capacity, and operational parameters including aircraft holding times, confidence levels, and energy consumption at the vertiport. Our findings reveal several critical insights. Basic M/M/1 queue models suggest that a server system's capacity can reach approximately 80\% of its total capacity while maintaining a reasonable queuing delay (around 5 minutes). However, due to interactions between the TLOF server and the surface server (parking pads), the actual throughput capacity of a vertiport is constrained around 50\% of its departure capacity for Markovian model. Moreover, confidence in the maximum aircraft holding time drops sharply once the arrival rate exceeds 40\% of the total departure capacity. We also find that determinism, which aims to reduce uncertainty in the aircraft holding time metric, is costly in terms of throughput; as we increase confidence in this metric, throughput decreases significantly. The simulation revealed that exponential parameters significantly reduce system capacity due to high associated variance. In our non-Markovian model, the system reaches full TLOF capacity, assuming there are adequate parking pads available. Additionally, we compared two-TLOF systems under different operational rules. Two independent TLOFs perform less effectively than separate TLOFs designated for takeoff and landing. Additionally, the throughput per unit area for two isolated TLOFs is lower due to smaller larger footprint requirement. Notably, no further increase in unit throughput is detected beyond 6 parking pads for the single TLOF configuration. These findings offer practical strategies for designing vertiport infrastructure to support the scalable, efficient, and sustainable deployment of urban air mobility systems.

\bibliographystyle{IEEEtran}
\bibliography{references}

@Article{drones7050319,
AUTHOR = {Al-Rubaye, Saba and Tsourdos, Antonios and Namuduri, Kamesh},
TITLE = {Advanced Air Mobility Operation and Infrastructure for Sustainable Connected eVTOL Vehicle},
JOURNAL = {Drones},
VOLUME = {7},
YEAR = {2023},
NUMBER = {5},
ARTICLE-NUMBER = {319},
URL = {https://www.mdpi.com/2504-446X/7/5/319},
ISSN = {2504-446X},
DOI = {10.3390/drones7050319}
}

@inproceedings{vascik2019development,
  title={Development of vertiport capacity envelopes and analysis of their sensitivity to topological and operational factors},
  author={Vascik, Parker D and Hansman, R John},
  booktitle={AIAA Scitech 2019 Forum},
  pages={0526},
  year={2019}
}

@article{scitech_vertisim,
author = {Emin Burak Onat and Vishwanath Bulusu and Anjan Chakrabarty and Mark Hansen and Raja Sengupta and Banavar Sridhar},
title = {Evaluating eVTOL Network Performance and Fleet Dynamics through Simulation-Based Analysis},
journal = {AIAA SCITECH 2024 Forum},
chapter = {},
pages = {},
year = {2024},
doi = {10.2514/6.2024-0336},
URL = {https://arc.aiaa.org/doi/abs/10.2514/6.2024-0336}
}

@article{onat2024simulationoptimization,
  title={A Simulation-Optimization Framework for Developing Wind-Resilient AAM Networks},
  author={Burak Onat, Emin and Cao, Shangqing and Rizwan, Raiyan and Jiang, Xuan and Hansen, Mark and Sengupta, Raja and Chakrabarty, Anjan},
  journal={arXiv e-prints},
  pages={arXiv--2405},
  year={2024}
}

@article{Newell1979,
author = {Newell, G. F.},
title = {Airport Capacity and Delays},
journal = {Transportation Science},
volume = {13},
number = {3},
pages = {201-241},
year = {1979},
doi = {10.1287/trsc.13.3.201},

URL = { 
https://doi.org/10.1287/trsc.13.3.201
},
eprint = { https://doi.org/10.1287/trsc.13.3.201
}
}

@misc{Malone1995,
	author = {Kerry Marie Malone},
	title = {Dynamic queueing systems: behavior and approximations for individual queues and for networks},
	year = {1995},
  note         = {Available at \url{https://dspace.mit.edu/handle/1721.1/11883}},
  school       = {Massachusetts Institute of Technology},
  type         = {PhD thesis}
}

@misc{Kivestu1976,
	author = {Peeter Andrus Kivestu},
	title = {Alternative methods of investigating the time dependent M/G/k queue},
	year = {1976},
  note         = {Available at \url{https://dspace.mit.edu/handle/1721.1/40340}},
  school       = {Massachusetts Institute of Technology},
  type         = {Masters thesis}
}

@article{simaiakis2016queuing,
  title={A queuing model of the airport departure process},
  author={Simaiakis, Ioannis and Balakrishnan, Hamsa},
  journal={Transportation Science},
  volume={50},
  number={1},
  pages={94--109},
  year={2016},
  publisher={INFORMS}
}

@article{Itoh2022,
  title = {Modeling Aircraft Departure at a Runway Using a Time-Varying Fluid Queue},
  volume = {9},
  ISSN = {2226-4310},
  url = {http://dx.doi.org/10.3390/aerospace9030119},
  DOI = {10.3390/aerospace9030119},
  number = {3},
  journal = {Aerospace},
  publisher = {MDPI AG},
  author = {Itoh,  Eri and Mitici,  Mihaela and Schultz,  Michael},
  year = {2022},
  month = feb,
  pages = {119}
}

@book{Devore_2012, 
place={Boston, MA}, 
edition={8}, 
title={Probability and statistics for engineering and the Sciences}, 
publisher={Brooks/Cole, Cengage Learning}, author={Devore, Jay L.}, 
year={2012}}

@book{Balsamo2001,
  title = {Analysis of Queueing Networks with Blocking},
  ISBN = {9781475733457},
  ISSN = {0884-8289},
  url = {http://dx.doi.org/10.1007/978-1-4757-3345-7},
  DOI = {10.1007/978-1-4757-3345-7},
  journal = {International Series in Operations Research &amp; Management Science},
  publisher = {Springer US},
  author = {Balsamo,  Simonetta and de Nitto Personé,  Vittoria and Onvural,  Raif},
  year = {2001}
}

@misc{SimPy, title={SimPy 4.0.2 Documentation},
 Year={2021},
 url={https://simpy.readthedocs.io/en/latest/}, journal={Overview - SimPy 4.0.2.dev1+g2973dbe documentation},
 note = {Accessed: 2024-03-24}
 }

@techreport{kim1995vertiport,
  title={Vertiport Capacity: Analysis Methods},
  author={Kim, Yeon-Myung and Schonfeld, Paul and Rakas, Jasenka},
  year={1995},
  institution={Federal Aviation Administration, Office of Communications, Navigation and Surveillance Systems.}
}

@article{zhang2022method,
  title={Method of Vertiport Capacity Assessment Based on Queuing Theory of Unmanned Aerial Vehicles},
  author={Zhang, Honghai and Fei, Yuhan and Li, Jingyu and Li, Bowen and Liu, Hao},
  journal={Sustainability},
  volume={15},
  number={1},
  pages={709},
  year={2022},
  publisher={MDPI}
}

@inproceedings{QueueEx1,
author = {Cooper, Robert B.},
title = {Queueing theory},
year = {1981},
isbn = {0897910494},
publisher = {Association for Computing Machinery},
booktitle = {Proceedings of the ACM '81 Conference},
pages = {119–122},
numpages = {4},
series = {ACM '81}
}

@misc{faa_brief,
title={Engineering Brief No. 105, Vertiport Design},
journal={Federal Aviation Administration},
howpublished = {\url{https://www.faa.gov/airports/engineering/engineering_briefs/engineering_brief_105_vertiport_design}},
year={2022}
}

@INPROCEEDINGS{UAMCapacity,
  author={Rimjha, Mihir and Trani, Antonio},
  booktitle={2021 Integrated Communications Navigation and Surveillance Conference (ICNS)}, 
  title={Urban Air Mobility: Factors Affecting Vertiport Capacity}, 
  year={2021},
  volume={},
  number={},
  pages={1-14},
  doi={10.1109/ICNS52807.2021.9441631}}

@article{balsamoreview,
title = {A review on queueing network models with finite capacity queues for software architectures performance prediction},
journal = {Performance Evaluation},
volume = {51},
number = {2},
pages = {269-288},
year = {2003},
note = {Queueing Networks with Blocking},
issn = {0166-5316},
author = {Simonetta Balsamo and Vittoria {De Nitto Personè} and Paola Inverardi},
}

@inproceedings{guerreiro2020capacity,
  title={Capacity and throughput of urban air mobility vertiports with a first-come, first-served vertiport scheduling algorithm},
  author={Guerreiro, Nelson M and Hagen, George E and Maddalon, Jeffrey M and Butler, Ricky W},
  booktitle={AIAA Aviation 2020 Forum},
  pages={2903},
  year={2020}
}

@INPROCEEDINGS{ShannonVertiportTopology,
  author={Zelinski, Shannon},
  booktitle={2020 AIAA/IEEE 39th Digital Avionics Systems Conference (DASC)}, 
  title={Operational Analysis of Vertiport Surface Topology}, 
  year={2020},
  volume={},
  number={},
  pages={1-10},
  doi={10.1109/DASC50938.2020.9256794}}

@techreport{heliportdesign,
  author = {{Federal Aviation Administration, U.S. Department of Transportation}},
  title = {Heliport Design Advisory Circular, AC No: 150/5390-2C},
  year = {2012},
  institution = {Federal Aviation Administration}}

@article{willemain2004,
title = {Statistical Analysis of Intervals between Projected Airport Arrivals},
author = {Thomas Willemain and Hui Fan and Huaiyu Ma},
number = {38-04-510},
year = {2004},
institution = {Rensselaer Polytechnic Institute}
}

@Article{gimpo_airport,
AUTHOR = {Ahn, Byeongseon and Hwang, Ho-Yon},
TITLE = {Design Criteria and Accommodating Capacity Analysis of Vertiports for Urban Air Mobility and Its Application at Gimpo Airport in Korea},
JOURNAL = {Applied Sciences},
VOLUME = {12},
YEAR = {2022},
NUMBER = {12},
ARTICLE-NUMBER = {6077},
URL = {https://www.mdpi.com/2076-3417/12/12/6077},
ISSN = {2076-3417},
DOI = {10.3390/app12126077}
}

@book{balsamoBook,
title = {Analysis of Queuing Networks with Blocking},
publisher={Springer},
author = {Balsamo, Simonetta},
year={2000}
}

@article{Erlang,
title = {The Theory of Probabilities and Telephone Conversations},
journal={Nyt Tidsskrift for Matematik},
author = {Agner Karup Erlang},
year={1909},
volume={B},
number={20}
}

@article{jackson1963jobshop,
  title={Jobshop-like queueing systems},
  author={Jackson, James R},
  journal={Management science},
  volume={10},
  number={1},
  pages={131--142},
  year={1963},
  publisher={INFORMS}
}

@techreport{wisk_conops,
author={Wisk and Skyports},
title={Concept of Operations: Autonomous UAM Aircraft Operations and Vertiport Integration},
series = {Technical Report},
year = {2022}}

@Inbook{Grottke2011,
author="Grottke, Michael
and Apte, Varsha
and Trivedi, Kishor S.
and Woolet, Steve",
editor="Boucherie, Richard J.
and van Dijk, Nico M.",
title="Response Time Distributions in Networks of Queues",
bookTitle="Queueing Networks: A Fundamental Approach",
year="2011",
publisher="Springer US",
address="Boston, MA",
pages="587--641",
isbn="978-1-4419-6472-4",
doi="10.1007/978-1-4419-6472-4_14",
url="https://doi.org/10.1007/978-1-4419-6472-4_14"
}

@misc{faa_vfr_rule,
  author       = {FAA},
  title        = {14 CFR 91.151(b) - General Operating and Flight Rules},
  howpublished = {\url{https://www.ecfr.gov/current/title-14/part-91#p-91.151(b)}}
}

@misc{faa_airport_profiles,
  author       = {FAA},
  title        = {Airport Capacity Profiles},
  howpublished = {\url{https://www.faa.gov/airports/planning_capacity/profiles}}
}

@misc{archerArcherAnnounces,
	author = {},
	title = {{A}rcher {A}nnounces {P}lanned {S}an {F}rancisco {A}ir {M}obility {N}etwork {C}onnecting {F}ive {L}ocations {A}cross the {B}ay {A}rea --- investors.archer.com},
	howpublished = {\url{https://investors.archer.com/news/news-details/2024/Archer-Announces-Planned-San-Francisco-Air-Mobility-Network-Connecting-Five-Locations-Across-the-Bay-Area/default.aspx}},
	year = {},
	note = {[Accessed 03-08-2024]},
}

@article{HATHERALL2024122848,
title = {Novel battery power capability assessment for improved eVTOL aircraft landing},
journal = {Applied Energy},
volume = {361},
pages = {122848},
year = {2024},
issn = {0306-2619},
doi = {https://doi.org/10.1016/j.apenergy.2024.122848},
url = {https://www.sciencedirect.com/science/article/pii/S0306261924002319},
author = {Ollie Hatherall and Anup Barai and Mona Faraji Niri and Zeyuan Wang and James Marco}
}

@article{Gomez2004,
title="Sojourn times in a two-stage queuing network with blocking",
author = "A. Gómez-Corral",
year=2004,
journal="Naval Research Logistics",
pages="1068--1089",
volume={51},
number={8}}

 

% \end{IEEEbiography}



% \vfill

\end{document}